\documentclass[aps,twocolumn,floats,prd,nofootinbib,showpacs,superscriptaddress,showkeys]{revtex4} %
\usepackage[dvips]{graphicx} %
\usepackage{epsfig,amsmath}
\usepackage{amssymb}
\usepackage{rotate}
\usepackage{color}
\usepackage{bm}
\usepackage{umoline}

\DeclareFontFamily{OT1}{pzc}{}
\DeclareFontShape{OT1}{pzc}{m}{it}%
            {<-> s * [1.10] pzcmi7t}{}
\DeclareMathAlphabet{\mathscr}{OT1}{pzc}%
                                {m}{it}

\newcommand{\be}{\begin{equation}}
\newcommand{\ee}{\end{equation}}
\newcommand{\bea}{\begin{eqnarray}}
\newcommand{\eea}{\end{eqnarray}}
\def\ba#1\ea{\begin{align}#1\end{align}}
       
\newcommand{\refeq}[1]{Eq.~(\ref{eq:#1})}          
\newcommand{\refeqs}[2]{Eqs.~(\ref{eq:#1})--(\ref{eq:#2})}          
          
\newcommand{\reffig}[1]{Fig.~\ref{fig:#1}}          
\newcommand{\reftab}[1]{Tab.~\ref{tab:#1}}          
          
\newcommand{\vs}{\nonumber\\}       
\newcommand{\refsec}[1]{Sec.~\ref{sec:#1}}          
\newcommand{\refapp}[1]{App.~\ref{app:#1}}          

\renewcommand{\v}[1]{\mathbf{#1}}

\newcommand{\vn}{\bm{\nabla}}
\newcommand{\vx}{\v{x}}

\newcommand{\vk}{\v{k}}

\newcommand{\vy}{\v{y}}

\renewcommand{\vr}{\v{r}}

\renewcommand{\binom}[2]{\left(\!\!\begin{array}{c}#1\\ #2 \end{array}\!\!\right)}

\newcommand{\<}{\langle}
\renewcommand{\>}{\rangle}

\newcommand{\nhat}{\hat{n}}

\newcommand{\eps}{\varepsilon}
\renewcommand{\d}{\delta}

\newcommand{\D}{\Delta}

\newcommand{\rhob}{\overline{\rho}}

\newcommand{\Mpch}{\,{\rm Mpc}/h}

\newcommand{\U}{\mathcal{U}}
\newcommand{\N}{\mathcal{N}}

\def\erfc{\mathrm{erfc}}

\newcommand{\s}{\sigma}

\newcommand{\M}{\mathcal{M}}

\renewcommand{\O}{\mathcal{O}}

\newcommand{\fNL}{f_{\rm NL}}

\def\L{\ell}

\def\Fh{F_{h,L}}

\begin{document}

\title{Peak-Background Split, Renormalization, and Galaxy Clustering}

\author{Fabian Schmidt}
\affiliation{Department of Astrophysical Sciences, Princeton University, Princeton, NJ 08544, USA}
\affiliation{Einstein fellow}

\author{Donghui Jeong}
\affiliation{
Department of Physics and Astronomy, Johns Hopkins University,
3400 N. Charles St., Baltimore, MD 21210, USA}

\author{Vincent Desjacques}
\affiliation{D\'epartement de Physique Th\'eorique and Center for Astroparticle Physics (CAP)
Universit\'e de Gen\`eve, 24 quai Ernest Ansermet, CH-1211 Gen\`eve, Switzerland}

\begin{abstract}
 We present a derivation of two-point correlations of general tracers in the 
 peak-background split (PBS) framework by way of a rigorous definition of 
 the PBS argument.  Our expressions only depend on connected matter 
 correlators and ``renormalized'' bias parameters with clear physical 
 interpretation, and are independent of any coarse-graining scale.  
 This result should be contrasted with the naive expression derived from 
 a local bias expansion of the tracer number density with 
 respect to the matter density perturbation $\d_L$ coarse-grained 
 on a scale $R_L$.  In the latter case, the predicted tracer correlation
 function receives contributions of order $\<\d_L^n\>$ at each perturbative 
 order $n$, whereas, in our formalism, 
 these are absorbed in the PBS bias parameters
 at all orders.  
 Further, this approach naturally predicts both a scale-dependent bias 
 $\propto k^2$ such as found for peaks of the density field, and
 the scale-dependent bias induced by primordial non-Gaussianity in the 
 initial conditions.  
 The only assumption made about the tracers is that their abundance at a 
 given position depends solely on the matter distribution within a finite 
 region around that position.  
\end{abstract}
\date{\today}



\maketitle

\section{Introduction}
\label{sec:intro}

The clustering of tracers of the large-scale structure (LSS) in the Universe,
such as galaxies, clusters, or the Lyman-$\alpha$ forest,
is one of the most important probes of the origin and evolution of
cosmological perturbations.  On sufficiently large scales, correlations
are weak and one should hope that a perturbative approach will allow us
to cleanly connect observations with the predictions of cosmological
models (such as the standard cold dark matter scenario with adiabatic Gaussian
initial conditions).  In case of the matter density perturbations, these
perturbative approaches are well developed (see \cite{Bernardeau/etal:2002}
for a review).  On the other hand, the formation of the tracers we actually
observe necessarily
involves highly non-linear, small-scale mechanisms which cannot be 
described perturbatively.  In order to relate observations to theory,
we thus need an effective description which involves unknown ``bias parameters''.  
These parameters in general need to be determined observationally, and
we would like as many parameters as necessary to accurately describe 
tracer correlations down to some minimum scale, but not
more, in order to retain the maximum amount of cosmological information.  
Which bias parameters need to be included, and to what order, is still
an open problem.  

The simplest and most well-known bias expansion is the local expansion
in terms of density \cite{szalay:1988,fry/gaztanaga:1983,coles:1993},
\be
\d_{h,L}(\vx) = c_0 + c_1 \d_L(\vx) + \frac{c_2}2 \d_L^2(\vx) + \dots,
\label{eq:localbias}
\ee
where $c_n$ are the bias parameters, $\d_{h,L}$ is the fractional tracer 
density perturbation, and $\d_L$ is the corresponding matter density 
perturbation.  Here, both the tracer and the matter density field are 
understood as coarse-grained on some scale 
$R_L$, so that this expansion can be interpreted as a counts-in-cells 
relation, and the $c_n$ as ``scatter-plot bias parameters''.  
This perturbative description is commonly assumed to be valid on large
scales \cite{scherrer/weinberg:1998,manera/gaztanaga:2011}, i.e. only 
if $\s_L^2 = \<\d_L^2\> \ll 1$, although the series could 
actually converge even for $\s_L\gtrsim 1$ if the condition 
$|c_{n+1}/c_n|\approx$~const. is satisfied in the limit $n\to\infty$;  however,
the larger $\s_L$, the more terms need to be included to obtain a converged
expression.  
As can easily be seen, the correlation function predicted by \refeq{localbias}
depends on the coarse-graining scale through the 
variance $\s_L^2 \equiv \<\d_L^2\>$, and all higher moments of the density field.  

In contrast to counts-in-cells studies \cite{fry/gaztanaga:1983}, where a specific scale $R_L$ is singled out, we expect that no additional smoothing scale $R_L$ should enter the calculation of correlation functions on a scale $r$ (unless we directly link $R_L$ to $r$, as done in \cite{chan/scoccimarro:2012}).  Thus, we need to
absorb the $R_L$-dependent pieces into
renormalized bias parameters $b_N$, as proposed for the first time by
\citet{mcdonald:2006} and tested against simulations in \citet{jeong/komatsu:2009a}.  More recently, this approach has been pursued to higher order in the multipoint propagator framework \cite{bernardeau/crocce/scoccimarro:2008} by \citet{matsubara:2011}.   In accordance
with renormalization theory, the expression for tracer correlators
in terms of the parameters $b_N$ should be $R_L$-independent, and convergent
in terms of matter correlators at the separation at which we measure
tracer correlations (rather than zero-lag correlators
such as $\<\d_L^2\>$).  In the process of renormalization, we have to introduce a new bias parameter $b_N$ at each order in $\d_L$ which needs to be determined from observations.    
Similar conclusions hold when adding other, non-local quantities to
the expansion \refeq{localbias}, such as derivatives of $\d$ or the tidal
tensor.  With in principle arbitrarily many free parameters, the
clustering of LSS tracers loses all cosmological constraining power.  
Moreover, treating the $b_N$ as mere nuisance parameters (as in 
\cite{mcdonald:2006,mcdonald/roy:2009}) precludes us from using any 
information contained in the $b_N$ on the parent halos and formation 
history of the tracers.

There is thus strong motivation to try to associate physical meaning with the
renormalized bias parameters.  This would allow models of, say, galaxy
formation to provide sensible priors on the allowed range of the
parameters, or to predict connections between different bias parameters.  For
example, there is a well-motivated connection between the linear
bias with respect to density of a tracer, and the parameter that
quantifies the scale-dependent bias induced by primordial non-Gaussianity
\cite{dalal/etal:2008,slosar/etal:2008}.  In this case, the connection
is crucial, as without it one would not be able to constrain
primordial non-Gaussianity from the scale-dependent bias.  

Our goal in this paper is to explicitly derive the physical meaning
of the renormalized bias parameters $b_N$, and their exact relation
to observables such as the correlation function to any order.  Physically, the renormalized bias parameters quantify the response of the mean abundance of tracers to a change in the background matter density $\rhob$ of the Universe (at fixed proper time since the Big Bang), i.e.
\be
b_N = \frac{\rhob^N}{\bar n}\frac{\partial^N \bar n}{\partial \rhob^N}\,.
\label{eq:bNintro}
\ee
This definition applies to tracers of any nature;  in the considerably more restrictive case of universal mass functions, where the abundance of tracers depends only on $\nu_c = \d_c / \s(M)$, a fractional change $D$ in the background density is equivalent to a change in the parameter $\d_c \to \d_c - D$, leading to the well-known peak-background split (PBS) bias parameters \cite{mo/jing/white:1997}, \refeq{bPBSth} in \refsec{buniv}.  Even though \refeq{bNintro} is much more general, we will still refer to the parameters $b_N$ as ``PBS biases'' for convenience, although it is important to keep the distinction in mind.  

We will show that the bias parameters defined through \refeq{bNintro}, and its generalization to non-local biases, are the coefficients multiplying powers of the matter correlation function $\xi_L(r)$ in the expansion of the tracer correlation function.  Speficially, for a Gaussian density field we obtain
\be
\xi_h(r) = \sum_{N=1}^\infty \frac{b^2_N}{N!} \left[\xi_L(r)\right]^N\,.
\ee
We also show that the $b_N$ agree with the bias parameters identified through a direct calculation of the  correlations of thresholded regions and peaks of the density field.  However, it is important to note the difference in philosophy between each derivation: thresholded regions and peaks constitute ``microscopic'' models of tracers, where the relation between matter and tracer density is explicitly specified on all scales (this is also the approach taken in \citet{matsubara:2011}, although the description is kept general there).  On the other hand, here we treat local biasing (and its generalization to non-local quantities) as effective description on sufficiently large scales, independently of the microscopic physics.  In the language of field theory, the former approaches constitute specific ``UV-complete'' theories, while the approach presented here is an ``effective theory'' of biasing.  Specifically, $R_L$ serves the role of a UV cutoff whose precise value should not impact correlations on scales of observable interest.

The exact relation between the parameters $b_N$ and tracer correlations provides a rigorous framework in which further modeling assumptions for any given tracer,
for example from the excursion set, peak model, or halo occupation
distribution, can be embedded --- both in order to tighten constraints
on cosmological parameters, and in order to infer the physics of the
formation of the tracers (as also pointed out in \cite{scoccimarro/etal:2012}, the deviations from the peak-background split predictions found in \cite{manera/etal:2010} are due to the fact that the authors assumed a universal
mass function of halos, not due to the inaccuracy of the peak-background split
argument itself).  Along the way, we also show
that renormalization removes the zero-lag matter correlators from the
expression for the tracer correlation function at all orders, as required.  
We also show that the same bias parameters describe both the tracer auto-
and the cross-correlation with matter.  The expression
of tracer correlations in terms of renormalized bias parameters and
``no-zero-lag'' matter correlators such as the correlation function 
is manifestly convergent as long as these matter correlators are small.  This is in close analogy to the expression for tracer correlations in terms of resummed bias parameters in \cite{matsubara:2011}.  Note that the treatment in \cite{matsubara:2011} is in Fourier space, while we work in real space here, for which we believe that the physical assumptions and arguments are more clear.

The renormalization approach proposed in \cite{mcdonald:2006} 
provides us with another extremely valuable tool:  
in describing the tracer density in terms of the ``bare'' bias parameters $c_n$,
we coarse-grain both tracer and matter fields on some scale $R_L$.  The
requirement that the final expression for observable tracer correlations
be independent of $R_L$ provides us with a quantitative estimate of the limits of
the ansatz \refeq{localbias}.  Once an $R_L$-dependence is found, we are
guided to find an additional dependence of the tracer density on a ``regional''
property of the matter density field 
which absorbs the dependence on $R_L$ through renormalization.  
We show this explicitly in 
two cases:  a bias with respect to $\nabla^2 \d$, previously found specifically for
peaks of the density field, which absorbs the $R_L$-dependence induced by the smoothing of the matter correlation function over the coarse-graining scale; and a bias with respect to the variance of
the small-scale density field, which has to be introduced in the presence
of primordial non-Gaussianity.  In the latter case, the renormalization
absorbs the $c_1 c_2 \<\d_L(1)\d_L^2(2)\>$ term into an overall 
scale-dependent bias coefficient, similarly as shown in \cite{mcdonald:2008}. 
However, in our approach we find a clear physical interpretation
of this coefficient, which preserves the connection to the Gaussian bias
derived in \cite{dalal/etal:2008,slosar/etal:2008}.  

In order to achieve these results efficiently, 
we adopt several simplifications.  First, we work
in real space rather than Fourier space, since this is where the
coarse-graining and ``separation of scales'' is defined.  We thus 
do not address effects such as stochasticity and exclusion, which are
restricted to very small physical separations (while they in general
affect Fourier-space correlations at all $k$).  Furthermore, our approach is primarily intended as being applied in Lagrangian space.  However, we will not restrict the treatment to a linear or Gaussian density field.  Thus, our results are applicable to biasing with respect to a nonlinearly evolved matter density field as well, though, in that case, further non-local biases should in general be considered \cite{desjacques/sheth:2010,chan/scoccimarro/sheth:2012,baldauf/etal:2012}.  We also assume that the tracer abundance
solely depends on the total matter distribution, rather than the baryon
and cold dark matter density separately;  the approach can easily be
generalized to deal with the two-component fluid.  
Finally, we restrict ourselves to two-point correlations of tracers.

Implicitly, we will work in synchronous-comoving gauge throughout.  That is, all
comoving observers on a constant-$t$ slice share the same proper time
(at linear order), and thus see a Universe of equal age.  For a different choice
of time slicing, tracer correlations in general receive contributions from
the different evolutionary stage of different regions as well \cite{gaugePk}
(see also \cite{yoo/etal:2009,challinor/lewis:2011,bonvin/durrer:2011,baldauf/etal:2011}).  

The outline of the paper is as follows.  In \refsec{PBS}, we describe
the basic approach assuming purely local biasing as in \refeq{localbias}
(although we never actually use this relation) and Gaussian initial
conditions, and derive the expression
for tracer correlations in terms of PBS bias parameters.  In \refsec{peakbias},
we show how a bias with respect to $\nabla^2\d$ naturally appears in this
approach.  
Finally, in \refsec{NG} we consider the case of non-Gaussian initial
conditions.  We conclude in \refsec{disc}.  The appendix contains the
derivation of the key result of \refsec{PBS} for general non-Gaussian
initial conditions, the extension of the treatment in \refsec{peakbias}
to higher order, and detailed derivations of some relations used in the
main text.  

\section{Peak-background split and tracer correlations}
\label{sec:PBS}

\begin{figure}[t]
\centering
\includegraphics[width=0.5\textwidth]{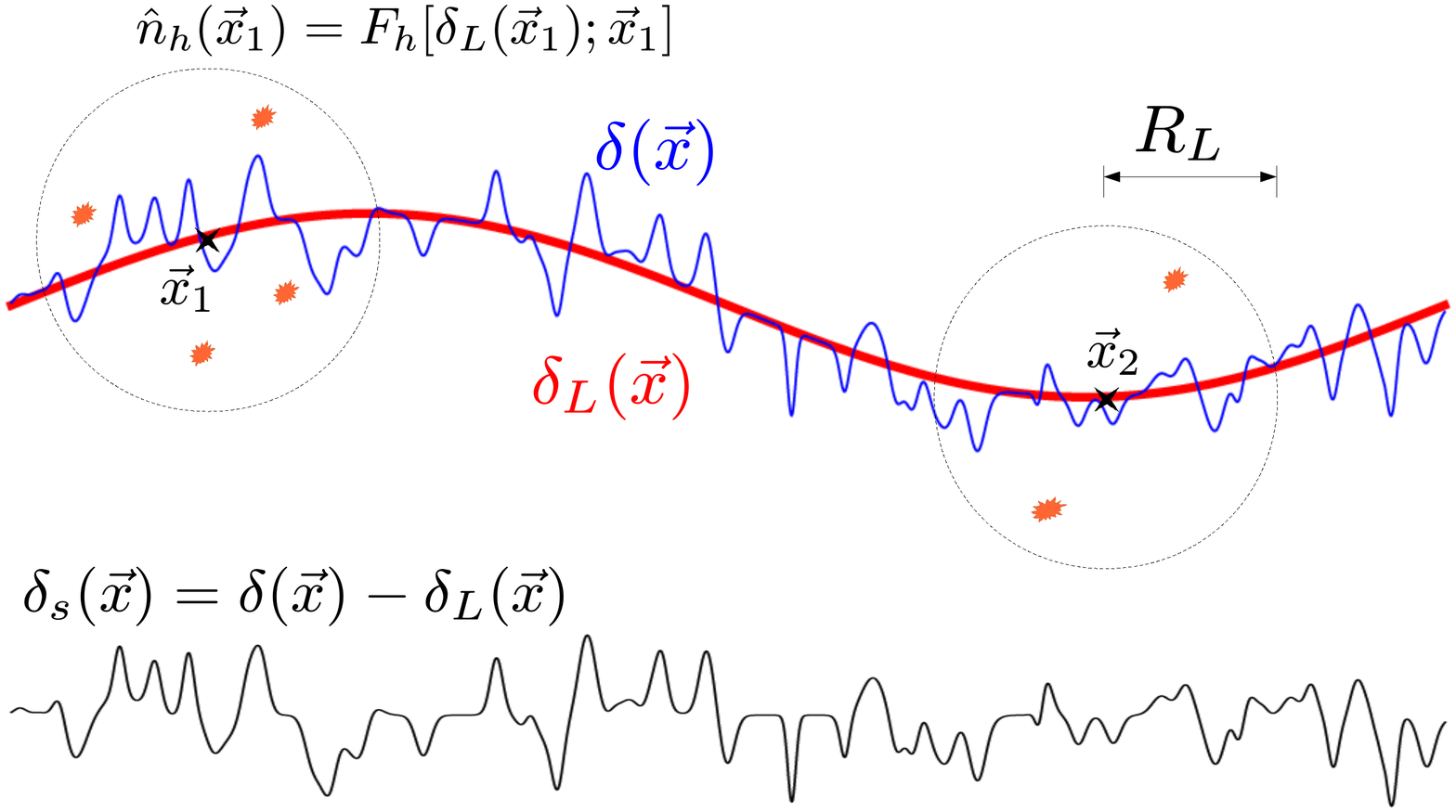}
\caption{Sketch of the separation of the density field (blue, thin line) 
into large-scale part $\d_L$ (red, thick line) and a small scale part $\d_s$
(\refeq{dsdef} in \refsec{NG}; thin black line below), via an arbitrary 
coarse-graining scale $R_L$.  
The tracer density coarse-grained on scale $R_L$ (circles) is described
by the function $\Fh(\d_L;\vx)$ [\refeq{Fp}], where the explicit dependence
on $\vx$ encodes the scatter around the mean relation with $\d_L$, which
is assumed to be uncorrelated with $\d_L$.
\label{fig:sketch}}
\end{figure}

Consider a filter function $W_L(\vx)$ of characteristic size $R_L$,
normalized to unity in 3D space.  In the following we assume that the filter 
function is isotropic,  $W_L = W_L(|\vx|)$.  This is a natural assumption 
since any anisotropy would correspond to introducing preferred directions.  
We define the filtered (coarse-grained) density field $\d_L$ in terms
of the full density field $\d(\vx)$ through
\ba
\d_L(\vx) \equiv\:& \int d^3\vy\:W_L(\vx-\vy) \d(\vy)\,,
\label{eq:dldef}
\ea
where the subscript $L$ refers to the coarse-graining scale $R_L$
(indicated as circles in \reffig{sketch}).  
We can think of $\d_L(\vx)$ as the average density within a region $\U$
of size $R_L$ centered on $\vx$.  
Note that while we primarily think of $\d(\vx)$ as being the Lagrangian
density field, many of our results will not make any assumptions about 
the statistics of $\d(\vx)$ (e.g., Gaussianity).  
The small-scale density field,
which we will consider in \refsec{NG}, is then defined as 
$\d_s(\vx) = \d(\vx) - \d_L(\vx)$ [\refeq{dsdef}].  
The number of tracers (orange dots in \reffig{sketch}) within this region 
$\U$ is simply given by the
weighted sum, i.e., the discretized analog of \refeq{dldef},
\be
\nhat_h(\vx) = \sum_i W_L(\vx_i-\vx),
\ee
where the sum runs over all tracers in the (idealized) sample, and
$\vx_i$ is the position of tracer $i$.  
The expectation value $\<\nhat_h\>$, estimated by averaging over $N$
different regions $\U$ and letting $N\rightarrow\infty$, is
equal to the cosmic mean of the
abundance of tracers.  It can be measured (with some uncertainty) 
either observationally or in simulations for any given tracer.  

We can now implicitly define a function $\Fh(\d_L;\vx)$ through
\be
\nhat_h(\vx) = \Fh(\d_L(\vx);\vx).  
\label{eq:Fp}
\ee
The dependence of $\Fh$ on $\vx$ denotes any departure, or ``scatter'',
of the tracer number density from a deterministic relation 
$\nhat_h(\vx) = \nhat_h[\d_L(\vx)]$;  by definition, this scatter is equivalent
to the dependence of $\nhat_h$ on the small-scale fluctuations $\d_s$ in the
given region.  The key assumption we will make
below is that the correlation of this scatter with large-scale perturbations
(in particular on the scales we are measuring correlation functions)
is negligible.  Then, the scatter will add noise to the measurement,
but will not contribute to the expectation value of correlation functions
on large scales \cite{dekel/lahav:1999}.  If this assumption breaks down, it
is straightforward 
to include other properties of the density field as arguments of $\Fh$,
which will be the subject of the following sections.  
The PBS argument we will apply below will allow us to derive the statistics
of the tracer without any explicit knowledge of the function $\Fh$.  As
indicated by the notation, the function $\Fh$ will depend on $R_L$.

We can formally expand \refeq{Fp} in  a Taylor series,
\be
\nhat_h(\vx) = \sum_{n=0}^\infty \frac1{n!} \Fh^{(n)}(0; \vx)\:[\d_L(\vx)]^n,
\label{eq:Fpseries}
\ee
where $\Fh^{(n)}(0;\vx)$ denotes the $n$-derivative of $\Fh(\d_L,\vx)$
with respect to $\d_L$ evaluated at position $\vx$ and at $\d_L=0$.  We now take the expectation value 
of \refeq{Fpseries} in order to obtain an expression for the mean number density 
of tracers, our first observable.  Our assumption of negligible correlation 
between $\d_L$ and the scatter encoded in the explicit $\vx$-dependence of $\Fh$ implies that the two factors in each term of \refeq{Fpseries} are 
independent random variables (cf. the Poisson clustering model \cite{coles:1993}):
\be
\< \Fh^{(n)}(0; \vx)\:[\d_L(\vx)]^n \> = \< \Fh^{(n)}(0; \vx)\> \:\< [\d_L(\vx)]^n \>
\ee
We then obtain
\ba
\<\nhat_h(\vx)\> &\:= \sum_n\frac{1}{n!}\left\<\Fh^{(n)}(0;\vx)\: [\d_L(\vx)]^n\right\>\\
&\:= \sum_n\frac{1}{n!}\left\<\Fh^{(n)}(0;\vx)\right\>\: \< \d_L^n\>\\
&\:= \<\Fh(0)\> \left( 1 + \frac{c_2}{2}\sigma_L^2 + \frac{c_3}{6} \<\d_L^3\>  + \dots \right),\label{eq:npavg}
\ea
where we have defined
\be
c_n \equiv \frac{1}{\<\Fh(0)\>} \left\<\Fh^{(n)}(0)\right\>,
\ee
dropping the argument $\vx$ since the ensemble average is independent of
location due to homogeneity.  The $c_n$ depend on $R_L$, but for clarity we will not indicate this dependence explicitly.  
Further, as in \refsec{intro},
\be
\s_L^2 \equiv \<\d_L^2\> = \int \frac{d^3k}{(2\pi)^3} |\tilde W_L(k)|^2 P(k).
\ee
Note that by definition, $\<\d_L\> = 0$, and $c_0=1$.    
In the limit that $R_L\rightarrow\infty$ so that $\s_L\rightarrow0$, we see 
that $\<\nhat_h\> = \<\Fh(0)\>$, the expectation value of the function
$\Fh$ at the background density.  For finite values of $R_L$ however, 
$\<\nhat_h\>$ receives contributions from the variance $\s_L^2$ and higher
order moments of the density field
coarse-grained with $W_L$.  This just says that for finite regions $\U$, 
$\<\Fh(0)\>$ does not give the cosmic mean of
the tracer abundance, $\<\nhat_h\>$.  
This is commonly phrased as a non-zero zeroth order bias parameter
given by $\<\nhat_h\>/\<\Fh(0)\>$ so that the cosmic mean is recovered 
upon taking the ensemble average.

\subsection{Correlations}

We now turn to the correlation function $\xi_h$ of tracers.   
If we measure the correlation function at separation $r$, we clearly need $R_L < r$ in
order to avoid large effects of the coarse graining.  However, as discussed 
in \refsec{intro}, the precise value of the coarse graining scale should not 
have an effect on the final expression for the correlation function.  
We will deal with the effects of the coarse-graining on the tracer 
correlation function in \refsec{peakbias}.

In terms of the coarse-grained densities $\nhat_h$, the simplest 
estimator for $\xi_h$ can be written as
\be
\hat\xi_h(r) = \sum_{r < |\vx_i-\vx_j| < r+\D r}\!\!\!\!
\frac{\nhat_h(\vx_i)\nhat_h(\vx_j)}{\bar n_h^2}  - 1.
\label{eq:xihat}
\ee
Here, $\bar n_h$ is the mean observed abundance of tracers, which can
be defined as $\bar n_h = N^{-1} \sum_i \nhat_h(\vx_i)$,
where the sum runs over a large number $N$ of random locations.  
We now let $\D r$ go to zero, and take the expectation value of
\refeq{xihat}:
\be
\<\hat\xi_h(r)\> = \frac{\<\nhat_h(1)\nhat_h(2)\>}{\<\nhat_h\>^2} - 1,
\label{eq:xih}
\ee
where `1' and `2' stand for two arbitrary locations separated by a 
distance $r$.  
Following the reasoning above, the derivatives of $\Fh$
with respect to $\d_L$ at the two locations separated by $r$ are
independent random variables,
\be
\< \Fh^{(n)}(0; \vx_1)\:\Fh^{(m)}(0; \vx_2) \> =
\< \Fh^{(n)}(0; \vx_1)\> \: \< \Fh^{(m)}(0; \vx_2) \> .
\ee
Using \refeq{Fpseries}, we can then write \refeq{xih} in terms
of the statistics of the coarse-grained density field $\d_L$, and the coefficients
$c_n$:
\be
\<\hat\xi_h(r)\> = \frac{\sum_{n,m=0}^\infty \frac{c_n c_m}{n! m!} \<\d_L^n(1)\d_L^m(2)\>}{\sum_{n,m=0}^\infty \frac{c_n c_m}{n! m!} \<\d_L^n\> \<\d_L^m\>} - 1.
\label{eq:xih2}
\ee
Similarly, we can define the tracer-matter cross-correlation function
(assuming that $\d_L$ is observable somehow)
and obtain its expectation value:
\ba
\hat\xi_{hm}(r) &\:= \sum_{r < |\vx_i-\vx_j| < r+\D r}\!\!\!\!
\frac{\nhat_h(\vx_i)\hat\d_L(\vx_j)}{\bar n_h}\\
\vspace*{0.2cm}
\<\hat\xi_{hm}(r)\> &\:= \frac{\<\nhat_h(1)\d_L(2)\>}{\<\nhat_h\>} \vs
&\:= \frac{\sum_{n=1}^\infty \frac{c_n}{n!} \<\d_L^n(1)\d_L(2)\>}{\sum_{n=0}^\infty \frac{c_n}{n!} \<\d_L^n\>}.\label{eq:xihm}
\ea
These expressions involve sums over moments of the density field, which contain
disconnected pieces such as $\<\d_L^2\>$, multiplied by the bare
bias parameters $c_n$.  The convergence of this sum of $R_L$-dependent coefficients
multiplying $R_L$-dependent disconnected moments is apparently only assured in
the general case if $\s_L^2$,
and higher moments of the density field, are much less than one, since
terms of order $\<\d_L^n\>$ appear at arbitrarily high $n$.  On the other hand,
a physically reasonable perturbative bias model for $\xi_h(r),\:\xi_{hm}(r)$ should converge as long 
as the \emph{connected} matter correlators are much less than one, independently of the
choice of the fictitious coarse-graining scale.  Thus, our goal is to reorder the
sum in \refeq{xih2} and \refeq{xihm} into a sum of $R_L$-independent coefficients
$b_N$ multiplying only powers of connected matter correlators.  This can be seen
as a renormalization of the ``bare'' coefficients $c_n$ into ``renormalized'' bias
parameters $b_N$.  We will
see below that they have clear physical significance.  

We can expand the correlators into connected parts (cumulants) as follows.  
The correlator appearing in the cross-correlation \refeq{xihm} can be written for any
statistical field as
\be
\<\d_L^n(1) \d_L(2) \> = 
\sum_{N=0}^n 
\binom{n}{N}
\<\d_L^{n-N}\> 
\<\d_L^N(1) \d_L(2) \>_c,
\label{eq:discon}
\ee
where the subscript $c$ denotes connected correlators.  Thus, \refeq{xihm}
can also be written as
\be
\<\hat\xi_{hm}(r)\> = \frac{1}{\N} \sum_{n=1}^\infty \frac{c_n}{n!} 
\sum_{N=0}^n 
\binom{n}{N}
\<\d_L^{n-N}\> \<\d_L^N(1) \d_L(2) \>_c,
\label{eq:xihm2}
\ee
where we have defined
\be
\N \equiv \sum_{n=0}^\infty \frac{c_n}{n!} \<\d_L^n\>.
\label{eq:N}
\ee
The density field correlator in the auto-correlation \refeq{xih2} is more 
complicated.  As shown in \refapp{general}, 
\be
\<\d_1^n \d_2^m\> = \sum_{k=0}^n\sum_{l=0}^m 
\binom{n}{k} \<\d_1^k\>
\binom{m}{l} \<\d_2^l\>
\<\d_1^{n-k} \d_2^{m-l}\>_{\rm nzl}\;,
\label{eq:dndm1}
\ee
where $\<\cdot\>_{\rm nzl}$ denotes a disconnected correlator that, when 
expanded into cumulants, does not contain any zero-lag pieces, i.e. no 
factors that asymptote to a constant as $r \to \infty$ (see \refeq{nzldef}
in \refapp{general} for a mathematical expression of this definition).  
We will restrict
to the Gaussian case here, and postpone the discussion of non-Gaussian density
fields to \refsec{NG} and \refapp{general}.  The no-zero-lag requirement
entails that $\d_L(1)$ and $\d_L(2)$ have to appear in equal powers in 
``nzl'' correlators.  Defining
\be
\xi_L(r) \equiv \<\d_L(1) \d_L(2)\>,
\ee
we obtain
\be
\<\d_L^n(1) \d_L^m(2)\>_{\rm nzl} \stackrel{\rm Gaussian}{=} n! \left[\xi_L(r)\right]^n\: \d_{nm}.
\label{eq:cnzlG}
\ee
The factor $n!$ comes about because there are $n!$ ways to contract 
$2n$ factors of $\d_L$ ($n$ of $\d_L(1)$ and $\d_L(2)$ each) into 
a product of $n$ correlation functions.  Further,
\be
\<\d_L^n\> = (n-1)!!\: \s_L^n
\label{eq:sRn}
\ee
for $n$ even, and zero for $n$ odd.  This yields
\ba
\<\d^n_L(1) \d^m_L(2)\>
\stackrel{\rm Gaussian}{=}\:& \hspace*{-0.5cm}\sum_{k:\  n-k,\:m-k\  \text{even}}^{{\rm min}(n,m)}\hspace*{-0.5cm}
(n-k-1)!! (m-k-1)!! \vs
&  \times
\binom{n}{k} \binom{m}{k}
k!\:
\sigma_L^{n+m-2k} [\xi_L(r)]^k.
\label{eq:dndm2}
\ea

\subsection{PBS and bias parameters}
\label{sec:PBSbias}

So far, all we have done is divide the universe into fictitious regions, 
and describe
the number density of tracers in regions in terms of a function $\Fh$
[\refeq{Fp}].  We were then able to formally express the correlations of tracers
in terms of the statistics of the matter density and the derivatives of
the function $\Fh$, all of which depend on the coarse-graining scale $R_L$.  

We now turn to the peak-background split argument, and the definition of the
PBS bias parameters with respect to density.  The argument can be summarized
as follows: if the description of the 
clustering of tracers solely through their dependence on $\d_L$ is sufficient, 
then the expected abundance of tracers in a region $\U$ characterized by
a coarse-grained overdensity $\d_L = D$ is sufficiently
well approximated by the average abundance of tracers $\<\nhat_h\>$ in a 
fictitious Universe with modified background density 
\be
\rhob' = \rhob (1+D)\,,
\label{eq:rhobprime}
\ee
where $\rhob$ is the actual background density.  

The advantage of this approach is that we only need a prediction for
$\<\nhat_h\>$ as function of the background density $\rhob'$ to calculate
the statistics of tracers;  no knowledge of the function $\Fh$ 
is necessary.  Note also that this is directly connected to the derivation
of bias (linear bias in that case) in the relativistic context
presented in \cite{gaugePk}.  Specifically, we are working in the 
synchronous gauge where all space-time
points on an equal-coordinate-time hypersurface share the same cosmic age.  
Correspondingly, when calculating $\<\nhat_h\>$ for varying $\rhob'$ it is
crucial to keep the age of the Universe fixed.  

Thus, we now consider the case where we perturb the background density
by $\D\rhob = D \rhob$, where $D$ is an infinitesimal parameter.  
Thus, in a region with overdensity $\d_L$ the matter density
is perturbed to
\be
\rho_L = \rhob (1 + \d_L) \to \rhob ( 1 + \d_L) + \D\rhob = 
\rhob ( 1 + \d_L + D)
\label{eq:rhobp}
\ee
Note that we add a fixed amount of \emph{uniform} matter density everywhere;
we do not rescale the local matter density $\rho$ by $1+D$, which
would also amplify the fluctuations $\d$.  
We can obtain the average number density of tracers $\<\hat n_h\>$ (more precisely,
the expectation value of the estimated mean number density in some volume) in
such a Universe from the expansion in terms of coarse-grained $\d_L$, \refeq{npavg}:
\be
\<\nhat_h\>|_D = \<\Fh(0)\> \sum_{n=0}^\infty \frac{c_n}{n!} \< (\d_L+D)^n \>,
\label{eq:npavgD}
\ee
where $\Fh$ and $c_n$ both refer to the Universe with background density
$\rhob$, i.e. $D=0$.  

Let us now define the \emph{peak-background split bias parameters} $b_N$
($N \geq 1$):
\be
b_N \equiv \frac{1}{\<\nhat_h\>|_{D=0}} \frac{\partial^N \<\nhat_h\>|_D}{\partial D^N}\Big|_{D=0}.
\label{eq:bN}
\ee
Using \refeq{rhobprime}, we can also write this as
\be
b_N = \frac{\rhob^N}{\<\nhat_h\>} \frac{\partial^N \<\nhat_h\>}{\partial \rhob^N}\,,
\ee
where the derivatives are evaluated at the fiducial value of $\rhob$.  
It is worth emphasizing the difference between these bias parameters and the
$c_n$ defined in the last section:  the $b_N$ quantify the response of
the cosmic mean abundance of tracers to a change in the background density
of the Universe; specifically, they do not make any reference to the regions
$\U$, or the scale $R_L$.  The $c_n$ on the other hand quantify the average 
response of the abundance of tracers within a region $\U$ to a change in the 
average density $\d_L$ within that region, evaluated at $\d_L=0$; they thus
necessarily depend on the filter function $W_L$ and scale $R_L$.  Further,
there is no ``scatter'' in the defining relation \refeq{bN} for the PBS 
biases, although the values for $b_N$ measured in reality
will clearly have a finite error bar as we can only approximate this 
relation within a finite volume.  

The $b_N$ are closely related to the resummed bias propagators defined in \cite{matsubara:2011} [see Eqs.~(83)--(84) there], while the bare bias parameters $c_n$ correspond to the bare propagators [Eqs.~(1)--(2) in that paper].  

Using \refeq{npavgD}, we can derive an algebraic relation between $b_N$
and $c_n$:
\ba
b_N &\:= \frac{1}{\N} \sum_{n=N}^\infty \frac{c_n}{n!} \frac{n!}{(n-N)!} 
\<\d_L^{n-N}\>.
\label{eq:bNcN}
\ea
By reordering the sum in \refeq{xihm2}, we immediately see that
the tracer-matter cross-correlation in terms of the PBS bias parameters
is given by
\be
\<\hat\xi_{hm}(r)\> = \sum_{N=1}^\infty \frac{b_N}{N!} \<\d_L^N(1) \d_L(2)\>_c.
\label{eq:xihmPBS}
\ee
This is the well-known bias expansion of the tracer-matter cross-correlation 
function.  However, note that the matter correlators appearing here
are the \emph{connected} correlators.  In particular, in the case of a 
Gaussian density field we obtain
\be
\<\hat\xi_{hm}(r)\> = b_1 \xi_L(r),
\label{eq:xihmPBSG}
\ee
i.e. the tracer-matter cross-correlation function is simply proportional to the linear
matter correlation function (the same conclusion was reached by \cite{frusciante/sheth:2012}, who only considered Gaussian density fields).  
Similarly, using \refeq{bNcN} and 
\refeq{dndm2} we can re-express the auto-correlation [\refeq{xih2}] as
\be
\<\hat \xi_h(r)\> = \sum_{N,M=1}^\infty \frac{b_N}{N!} \frac{b_M}{M!}
\<\d_L^N(1) \d_L^M(2)\>_{\rm nzl}\,.
\label{eq:xihPBS}
\ee
This relation straightforwardly generalizes to the cross-correlation
between two different tracers $h_1,\,h_2$, yielding
\be
\<\hat \xi_{h_1 h_2}(r)\> = \sum_{N,M=1}^\infty \frac{b^{(1)}_N}{N!} \frac{b^{(2)}_M}{M!}
\<\d_L^N(1) \d_L^M(2)\>_{\rm nzl}\,,
\nonumber
\ee
where $b_N^{(i)}$ denotes the PBS bias prameter for tracer $h_i$ [\refeq{xihmPBS} is of course a special case of this, with $b_1=1,\, b_{N>1} =0$ for matter].

Even though \refeq{dndm2} assumes a Gaussian density field, \refeq{xihPBS}
is in fact valid for a general non-Gaussian density field.  
The proof in this case requires somewhat more effort and is given in \refapp{general}.  We point out that the derivation of \refeq{xihPBS} is equivalent to the renormalization of multi-point
propagators \cite{bernardeau/crocce/scoccimarro:2008,bernardeau/crocce/sefusatti:2010,matsubara:2011}, and valid for general statistical fields.  
Note that there is no $b_0$;  the expressions \refeqs{xihmPBS}{xihPBS} only involve terms with $b_N$ for $N \geq 1$.  
In the Gaussian case, \refeq{xihPBS} further simplifies to
\be
\<\hat \xi_h(r)\> = \sum_{N=1}^\infty \frac{b^2_N}{N!} \left[\xi_L(r)\right]^N.
\label{eq:xihPBSG}
\ee
\refeqs{xihmPBS}{xihPBSG} achieve the desired result: an expansion of the
tracer correlation function in terms of $R_L$-independent bias parameters
which multiply powers of the matter correlation function $\xi_L(r)$ (or, more
generally, no-zero-lag correlators).  The series in \refeqs{xihPBS}{xihPBSG} 
have a convergence radius set solely by the values of the $b_N$ and the amplitude
of the matter correlation function at scale $r$, which is what we expect from a physical bias expansion.\footnote{If $b_N^2/N!$ grows faster with $N$ than an exponential, e.g, if $b_N \sim N^{a N}$ with $a > 1/2$, then the series does not converge for any $r$.  In this case, our approach does not make a prediction for the clustering of tracers in the large-scale limit.  Through \refeq{bN}, this essentially means that $\<\hat n_h\>$ is defined only for one exact value of $\rhob$ and nowhere else, which is clearly not a physical behavior.} On the other hand, in the
bare bias expansion, \refeq{xih2}, terms of order $\sigma_L^n$ appear at every
successive higher order, suggesting that we need to coarse-grain the density
field on quasi-linear scales in order for the expansion to be perturbatively valid.  This
of course would be disastrous for any sharp features in the correlation
function; for example, choosing a coarse-graining scale of $R_L=30-50\Mpch$ would
erase the baryon acoustic oscillation (BAO) feature at $r=150\Mpch$.  In the PBS
bias expansion, there is no need to choose a coarse-graining scale this large.  
Rather, the validity of the the result \refeqs{xihmPBS}{xihPBSG} is determined
by the requirement that any residual dependence on $R_L$ be negligible.  

Another important property of this expansion is that the same PBS bias 
parameters describe both the matter-tracer cross-correlation
and the tracer auto-correlation, which is what we expect from a physical 
bias expansion (the corresponding
statement in Fourier space is complicated by the small-scale effects such as
stochasticity, shot noise and halo exclusion
which contribute to the tracer power spectrum at all $k$).

One crucial advantage of this approach is that we have an indicator for when
the underlying assumptions break down: if evaluation of \refeq{xihPBS} shows
that the result is in fact $R_L$-dependent, then we know that the underlying
assumptions, in particular the description $\hat n_h(\vx) = \Fh(\d_L(\vx))$, break down.  One then has to identify the physical reason for this $R_L$-dependence, and add a dependence of the tracer number density on additional parameters which will absorb (``renormalize'')
the $R_L$-dependence.  We will see two important examples of this in 
\refsec{peakbias} and \refsec{NG}.  First however, we will derive the PBS biases for the widely considered case of universal mass functions, and then illustrate the
approach on a concrete example of biased tracer: regions above threshold.

\subsection{PBS biases for universal mass functions}
\label{sec:buniv}

The mean abundance of tracers such as dark matter halos of some mass $M_*$ is often
parametrized in the form 
\ba
\bar n_h =\:& \rhob\:f(\nu_c)\: J_* 
\label{eq:univ1}\\
\nu_c \equiv\:& \frac{\d_c}{\s_*}; \quad J_* \equiv \frac{d\ln \s_*}{d\ln R_*},
\label{eq:Js}
\ea
where $\s_*$ is the variance of the linear matter density field on scale $R_*$, $R_*$ is related to the mass $M_*$ through $M_* = 4\pi/3\, \rhob R_*^3$, and $\d_c$ is the linearly extrapolated threshold for collapse.  Further, $f(\nu_c)$ is in general an arbitrary function of $\nu_c$.  The Jacobian $J_*$ is present in order to convert from an interval in $\s_*$ to a mass interval.  \refeq{univ1} is referred to as ``universal mass function'' and was originally motivated by the excursion set formalism \cite{press/schechter:1974}.  It is a special case of a more general description of mean tracer abundance we will consider in \refsec{univ}.   

In order to derive the bias parameters \refeq{bN}, we need to know how $\bar n_h$ changes under a change in the background density of the Universe [\refeq{rhobprime}].  Since we work in the Lagrangian picture, we will ignore the trivial dependence through the $\rhob$ prefactor in \refeq{univ1}.  The threshold $\d_c$ is defined as the fractional overdensity a region must have to collapse\footnote{Since General Relativity is scale-free, this threshold is independent of the size and enclosed mass of the perturbation.} at a fixed proper time $t_0$.  In an Einstein-de Sitter Universe, a spherical perturbation with a mean initial fractional overdensity
$\d_c \approx 1.686$, i.e. with $\rho(<R,t) = \left[1 + a(t) \d_c\right] \rhob(t)$ average interior density for $a(t) \ll 1$, collapses at $a=1$.  The same reasoning also holds for more general expansion histories, where $\d_c$ assumes other values.  Since the evolution of such a perturbation is independent of the external Universe (by Birkhoff's theorem), a perturbation of the same physical density $\rho_c$ will collapse at the same proper time in a Universe with perturbed background density $\rhob' = \rhob(1+D)$ as well.  The significance $\nu_c = \d_c/\s_* = (\rho_c-\rhob) / \d\rho_{\rm RMS}$ quantifies how rare fluctuations above a physical density threshold $\rho_c= (1+\d_c)\rhob$ are given the RMS fluctuation amplitude $\d\rho_{\rm RMS} = \s_* \rhob$.  Clearly, if we add a uniform matter density component $D\,\rhob$, the critical overdensity changes to
\be
\rho_c - \rhob' = (1 + \d_c)\rhob - (1 + D)\rhob = (\d_c - D) \rhob\,.
\ee
Thus, the significance is modified to
\be
\nu_c' = \frac{\rho_c-\rhob'}{\s_*\rhob} = \frac{\d_c - D}{\s_*}\,.
\ee
For a mass function of the form \refeq{univ1}, changing the background density is thus equivalent to changing $\d_c \to \d_c - D$.  \refeq{univ1} and \refeq{bN} thus immediately yield
\be
b_N = \frac{(-1)^N}{\<\nhat_h\>}\frac{\partial^N\<\nhat_h\>}{\partial \d_c^N}
= \frac{(-1)^N}{\sigma_*^N} \frac{1}{f(\nu_c)}
\frac{d^N f(\nu_c)}{d \nu_c^N}.
\label{eq:bPBSuniv}
\ee
This is the widely known expression for the peak-background split bias parameters, which in our approach is a special case of \refeq{bN}.

\subsection{Application to regions above threshold}
\label{sec:ex}

We now turn to a simple example of tracer for which an \emph{exact} expression
of the tracer correlations is known.  Precisely, we define our
tracer to be a region where the density field $\d$ is above a fixed threshold
$\d_c = \nu_c \s_*$, where $\s_*$ is the RMS fluctuation of the density field
and $\nu_c$ an arbitrary fixed parameter.  The density field can be thought
of as smoothed on some scale $r_*$; however, since for our purposes this 
scale is irrelevant, we will not make this smoothing explicit in our notation
in order to avoid confusion.  Note also that we do not make any assumption
about $\nu_c$, such as the ``high peak limit''. 
In the present case, unlike peaks of the density field, 
the tracer population is not a (countable) point set.  
Rather, tracers cover a finite volume, and the tracer number 
can be defined as a continuous field 
\be
N_h(\vx) = \Theta(\d(\vx) - \nu_c\sigma_*).
\label{eq:nh}
\ee
The number density of tracers as defined earlier in this section is then given by
\be
\nhat_h(\vx) = \int d^3\vy N_h(\vy) W_L(\vx-\vy).
\ee  
We first review the exact approach to the clustering of such ``tracers'', and then 
investigate the PBS prediction.  Note that essentially all these results 
have already been derived in \cite{kaiser:1984,matarrese/etal:1986,jensen/szalay:1986,
matsubara:1995,long,matsubara:2011,ferraro/etal:2012}.
However, we review it here in light of the discussion presented above.

\subsubsection{Exact calculation for a Gaussian density field}
\label{sec:exact}

The mean ``number density'' $\<\nhat_h\>$ of the tracers defined above
is simply given by the fraction of the total volume that is above the
threshold $\nu_c \s_*$,
\be
\<\nhat_h\> = P_1(\nu_c),
\label{eq:npex}
\ee
where $P_1(\nu_c)$ is the probability that the density
field at a random location is larger than $\nu_c\sigma_*$.  
If the underlying density field follows Gaussian statistics 
with variance $\sigma_*$, $P_1$ is given by
\be
P_1(\nu_c) =
\frac{1}{\sqrt{2\pi}}
\int_{\nu_c}^\infty dx e^{-x^2/2} 
= \frac{1}{2}\mathrm{erfc}\left(\frac{\nu_c}{\sqrt{2}}\right).
\label{eq:P1}
\ee
The exact expression for the two-point correlation function of our ``tracers'',
$\xi_h(r)$, is then given by 
the probability of finding two peaks $P_2$ separated by $r$, relative
to the random probability (Sec.~III~B of \cite{long}) through
\ba
\xi_h(r) =\:& \frac{P_2(\nu_c;r)}{[P_1(\nu_c)]^2} -1 \vs
=\:& \frac{2}{\pi}\left[\erfc\left(\frac{\nu_c}{\sqrt{2}}\right)\right]^{-2}\vs
&\times \sum_{N=1}^\infty
\frac{\bigl[\xi(r)\bigr]^N}{N!\sigma_*^{2N}}
\bigl[H_{N-1}(\nu_c)\bigr]^2
e^{-\nu_c^2}.
\label{eq:2ptfn_G1}
\ea
Here, $\xi(r)$ is the two-point
correlation function of the underlying density field smoothed on the scale $R_*$, 
so that $\xi(0)=\s_*^2$.  

\subsubsection{Peak-background split calculation}

\refeq{xihPBS} gives the tracer correlation function in terms of
the PBS bias parameters and powers of $\xi_L(r)$.  
Note however that given the explicit relation between $\nhat_h$ and $\d$
through \refeq{nh} in this simple example, the division into
regions $\U$ is a purely conceptual device here, and we can always
set $R_L$ to be equal to or smaller
than the smoothing scale adopted in the thresholding
approach.  Hence, we will drop the subscript $R_L$ below.  

The PBS bias parameters are
defined by \refeq{bN}.  Given our definition of tracers as regions where
$\d(\vx) > \d_c$, or equivalently, $\rho(\vx) > \rhob(1+\d_c)$, we see
that a fractional change $D$ in the background density is
equivalent to a change in the threshold $\d_c$:
\be
\rhob \rightarrow \rhob(1+D)\; \Leftrightarrow \;\d_c \rightarrow \d_c-D.
\ee
Hence, the PBS bias parameters are given by
\be
b_N = \frac{(-1)^N}{\<\nhat_h\>}\frac{\partial^N\<\nhat_h\>}{\partial \d_c^N}
= \frac{(-1)^N}{\sigma_*^N} \frac{1}{P_1(\nu_c)}
\frac{d^N P_1}{d\nu_c^N}.
\label{eq:bPBSth}
\ee
This of course can also be derived by noting that the ``abundance'' of regions above treshold \refeq{npex} is a special case of universal mass functions (\refeq{univ1} without the Jacobian factor which is irrelevant for the $b_N$), so that \refeq{bPBSuniv} applies.  
By using the generating function of the (probabilists') Hermite polynomial
\begin{equation}
H_n(x) = (-1)^n e^{x^2/2} \frac{d^n}{dx^n}
\left(
e^{-x^2/2}
\right)
\end{equation}
we calculate the $n$-th derivative of $P_1$ ($n \geq 1$) as 
\begin{align}
\nonumber
\frac{d^nP_1}{d\nu_c^n}
=\:&
\frac{d^{(n-1)}}{d\nu_c^{(n-1)}}
\left(
-
\frac{e^{-\nu_c^2/2}}{\sqrt{2\pi}}
\right)
\\
=\:&
(-1)^n H_{n-1}(\nu_c)
\frac{e^{-\nu_c^2/2}}{\sqrt{2\pi}}.  
\end{align}
Thus, we can explicitly write the PBS bias parameters for our tracers:
\begin{align}
b_N =\:& \sqrt{\frac2\pi} \left[
\erfc\left(\frac{\nu_c}{\sqrt{2}}\right)\right]^{-1}
\frac{e^{-\nu_c^2/2}}{\sigma_*^N}
H_{N-1}(\nu_c).
\label{eq:bPBSthr}
\end{align}
Inserting this into \refeq{xihPBSG}, we immediately obtain the PBS
prediction for the correlation of thresholded regions:
\ba
\xi_h(r) =\:& \frac2\pi \left[
\erfc\left(\frac{\nu_c}{\sqrt{2}}\right)\right]^{-2}\!\!\! e^{-\nu_c^2}
\sum_{N=1}^\infty \frac{H_{N-1}(\nu_c)}{N!\,\sigma_*^{2N}} \left[\xi(r)\right]^N.
\ea
We see that this agrees with the direct calculation, \refeq{2ptfn_G1}.  
A mathematically similar derivation was presented in \cite{long}.  The
difference is that in \cite{long}, we used \refeq{bPBSth} and
\refeq{2ptfn_G1} to \emph{infer} the general relation \refeq{xihPBSG}.  
Here, we are simply illustrating how the independently derived \refeq{xihPBSG} applies
to the case of thresholded regions, a case where we know explicitly
the function $\Fh(\d_L)$.  As proven in the previous section,
\refeq{xihPBSG} and the much more
general \refeq{xihPBS} apply to \emph{any} tracer as long as
the dependence of the tracer density on other quantities apart from
the matter density can be neglected.  

One alternative to the simple local bias expansion \refeq{localbias}
in the context of thresholded regions is to expand \refeq{nh} in terms of
Hermite polynomials, as done in \cite{matsubara:1995,matsubara:2011,ferraro/etal:2012} (see also \cite{szalay:1988,coles:1993}):
\ba
\Theta(\nu-\nu_c) = 
\sum_{n=0}^\infty a_n(\nu_c)H_n(\nu)
\ea
where $\nu(\vx) \equiv \d(\vx)/\s_*$, and
\be
a_n(\nu_c) = \left\{
\begin{array}{ll}
\frac12 {\rm erfc}\left(\frac{1}{\sqrt{2}}\nu_c\right) & n=0\\
\frac1{n!} \frac{1}{\sqrt{2\pi}}e^{-\nu_c^2/2}H_{n-1}(\nu_c) & n\ge1
\end{array}.
\right.
\ee
The bias parameters they obtain are exactly equal
to our renormalized PBS bias parameters, \refeq{bPBSthr}.  The reason
for this is that, in the Gaussian case, the Hermite expansion ensures that
no disconnected pieces remain in the correlation function expression
\refeq{xih2}.  Specifically, denoting $\nu_i = \nu(\vx_i)$, 
we have 
\ba
& \< H_n(\nu_1) H_m(\nu_2) \> = \frac1{2\pi} \int_{-\infty}^\infty d\nu_1\:H_n(\nu_1)
\int_{-\infty}^\infty d\nu_2\:H_m(\nu_2) \vs
& \hspace*{3.2cm} \times \exp\left(\frac{\xi(r_{12})}{\s_*^2}
\frac{\partial^2}{\partial \nu_1\partial \nu_2}\right) e^{-(\nu_1^2+\nu_2^2)/2} \vs
& = \frac1{2\pi}\sum_{N=0}^\infty \frac{1}{N!}\left(\frac{\xi(r_{12})}{\s_*^2}\right)^N \int d\nu_1 H_n(\nu_1) H_N(\nu_1) e^{-\nu_1^2/2} \vs
& \qquad \times \int d\nu_2 H_m(\nu_2) H_N(\nu_2) e^{-\nu_2^2/2} \vs
& = n! \left(\frac{\xi(r_{12})}{\s_*^2}\right)^n \d_{nm}.
\ea
The Hermite expansion is thus an elegant way of directly obtaining
renormalized bias parameters in the case of thresholding in a Gaussian
density field.  However, the additional contributions obtained in the 
non-Gaussian case spoil this renormalization of all zero-lag terms, 
as we will see in \refsec{NG}.  

\section{Smoothed correlation function and curvature bias}
\label{sec:peakbias}

Above we explained that the expression of the tracer correlation in terms
of PBS bias parameters and connected matter correlators  
[\refeq{xihPBS} and \refeq{xihPBSG} for the general and Gaussian case,
respectively] 
should be numerically insensitive to the value of the coarse-graining 
scale $R_L$.  Further, a significant $R_L$-dependence indicates the break-down 
of our assumption that the tracer density is a function only of the local 
coarse-grained density.  

For $r \gg R_L$, and as long as $\xi(r)$ is smooth (e.g., a power law), 
the smoothed version of $\xi(r)$, $\xi_L(r)$, will not differ significantly
from $\xi(r)$.  However, if $\xi(r)$ has some features on a scale $\d r \ll r$, 
such as the BAO feature with $\d r \sim 20 \Mpch$, 
then the condition for $R_L$-independence becomes much more 
restrictive: $R_L \ll \d r$.   We now show how the $R_L$-dependence induced
through $\xi_L(r)$ can be cured.  

For a general
isotropic filter function (see \refeq{Wf} in \refapp{peakbias})
the effect of smoothing on the correlation function $\xi(r)$ can be
perturbatively described through
\ba
\xi_L(r) =\:& \int\frac{d^3k}{(2\pi)^3} |\tilde W_L(k)|^2 P(k) e^{i\vk \cdot\vr} \vs
=\:& \int\frac{d^3k}{(2\pi)^3} \left(1 - 2 R_L^2 k^2 + \O(k^4)\right) P(k) e^{i\vk \cdot \vr} \vs
=\:& \xi(r) + 2 R_L^2 \nabla^2 \xi(r) + \O(\nabla^4 \xi(r)),
\label{eq:xiRexp1}
\ea
by suitable definition of the parameter $R_L$.  In \refapp{peakbias},
we give the general expansion of $\tilde W_L(k)$ [\refeq{Wtgen}]
and the expansion of $\xi_L(r)$ in terms of derivatives of $\xi(r)$
[\refeq{xiRexp}].  
Thus, if $R_L^2 \nabla^2 \xi$ is comparable to $\xi$, our requirement of $R_L$-independence does not
hold.  In our approach, this signals a breakdown of the underlying assumption
that tracer statistics can be described purely by their dependence on the local
matter density.  Instead, let us assume that the local number density also
depends on the coarse-grained Laplacian of the density field, i.e. the
curvature:
\be
\nhat_h(\vx) = \Fh(\d_L(\vx); \nabla^2\d_L(\vx); \vx).  
\label{eq:Fpn}
\ee
The Laplacian is the lowest order term in derivatives of $\d_L$, because a
dependence on the gradient of $\d_L$ would imply a preferred direction\footnote{Terms such as $(\vn\d_L)^2$ could however appear at second and higher order.}.  
In general, we now have to perform a bivariate expansion of the function
$\Fh$ in $\d_L$ and $\nabla^2 \d_L$.  Let us for now restrict to lowest order
to keep the treatment clear, and consider the Gaussian case.  The expansion
to higher orders in derivatives of $\d_L$ is described in \refapp{peakbias}.  
The tracer auto-correlation becomes
\ba
\xi_h(r) =\:& c_1^2 \<\d_L(1) \d_L(2)\> + 2 c_1 c_{\nabla^2 \d} \<\d_L(1) \nabla^2\d_L(2)\> \vs
& + \O(\nabla^4 \xi) \vs
=\:& c_1^2 \left[\xi(r) + 2 R_L^2 \nabla^2 \xi(r)\right] + 2 c_1 c_{\nabla^2\d} \nabla^2 \xi(r).
\label{eq:xihn1}
\ea
Here, we have defined
\be
c_{\nabla^2\d} = \frac{1}{\<\Fh(0)\>} \left\<\frac{\partial \Fh}{\partial(\nabla^2\d_L)}\Big|_{\d_L=0,\nabla^2\d_L=0}\right\>.
\label{eq:cnabdef}
\ee
\refeq{xihn1} is again phrased in terms of (in general) disconnected matter
correlators and $R_L$-dependent bare bias parameters.  We now need to introduce
a $R_L$-independent PBS bias parameter for $\nabla^2 \d$ as defined in \refsec{PBSbias} 
for the density itself.  We would like a transformation where the Laplacian
of the density perturbation shifts by a constant:
\be
\nabla^2\d_{\alpha}(\vx) = \nabla^2\d(\vx) + \frac{\alpha}{\L^2},
\label{eq:curvtransf}
\ee
where $\alpha$ is a dimensionless small parameter, and we have added a length 
scale $\L$.  This corresponds to
\be
\d_{\alpha}(\vx) = \d(\vx) + \frac{\alpha}{6\L^2} \left(\vx^2 + \v{A}\cdot \vx + C \right),
\label{eq:alphatransf1}
\ee
where $\v{A}$ and $C$ are constants and the center of the region considered is
chosen as the origin.  
We are not interested in adding a gradient to
the density field and hence set $\v{A} = 0$.  
Note that \refeq{alphatransf1} is only defined for a 
region of finite size (e.g., a simulation box), so that $\<n_h\>$ in the following 
is to be considered as an ensemble average over many such finite regions.  We will
also set $C=0$ so that $\d(\v{0})$, at the center of the region considered, is
unchanged (a constant shift in $\d$ such as described by $C$ just corresponds to
the density bias transformation of \refsec{PBSbias}).  Thus,
\be
\d_\alpha(\vx) = \d(\vx) + \frac{\alpha}{6\L^2} \vx^2.
\label{eq:alphatransf}
\ee
We can now defined a (renormalized) PBS bias parameter through
\be
b_{\nabla^2\d} = \frac{\L^2}{\<\nhat_h\>} \frac{\partial \<\nhat_h(\v{0})\>}{\partial\alpha}\Big|_{\alpha=0}.
\label{eq:bndef}
\ee
Defined in this way, the scale $\L$ will disappear out of the final expression
for the tracer correlation function (note that $b_{\nabla^2\d}$ has dimension
length squared).  
In order to derive the relation between $b_{\nabla^2\d}$ and the $c_n$, we need the
behavior of both $\d_L(\vx)$ and $\nabla^2 \d_L(\vx)$ under the transformation
\refeq{alphatransf}.  The latter is immediately obtained as
\be
\nabla^2 \d_{L,\alpha}(\vx) = \nabla^2 \d_L(\vx) + \frac{\alpha}{\L^2}.
\ee
We are interested in the change of $\d_L$ near the origin (the center 
of the region $\U$).  In analogy with \refeq{xiRexp1}, we obtain
\ba
\d_{L,\alpha}(\v{0}) =\:& \int d^3\vy\: W_L(\vy) \left[\d(\vy) + \frac16 \frac{\alpha}{\L^2} \vy^2 \right] \vs
=\:& \d_L(\v{0}) + \alpha \frac{R_L^2}{\L^2}.
\ea
Thus, \refeq{Fpn} yields at lowest order
\ba
b_{\nabla^2 \d} & = \frac{\L^2}{\Fh(0)}
\vs
&\times\left(\frac{\partial \Fh(\d_L,0)}{\partial\d_L} \frac{\partial\d_L}{\partial \alpha}  
+ \frac{\partial \Fh(0,\nabla^2\d_L)}{\partial(\nabla^2\d_L)} \frac{\partial(\nabla^2\d_L)}{\partial \alpha} \right)\vs
& = c_1 R_L^2 + c_{\nabla^2 \d}.
\label{eq:bn1cn1}
\ea
Now, we can write the generalization of our previous result, \refeq{xihPBSG}, at lowest
order, assuming we absorb the smoothing effect on $\xi(r)$, as
\ba
\xi_h(r) = b_1^2 \xi(r) + 2 b_1 b_{\nabla^2\d} \nabla^2 \xi(r),
\label{eq:xihnPBS}
\ea
where the factor of 2 comes from the two permutations when writing down all mixed
no-zero-lag correlators between $\d$ and $\nabla^2\d$.  Using \refeq{bn1cn1}, this
is equal to
\ba
\xi_h(r) = c_1^2 \xi(r) + 2 \left[c_1^2 R_L^2 + c_1 c_{\nabla^2 \d} \right] \nabla^2\xi(r) + \O(\nabla^4\xi(r)),
\nonumber
\ea
exactly matching the result of \refeq{xihn1}.  Thus, by introducing a dependence
of the tracer density on the Laplacian of the density field, and a corresponding
PBS bias parameter, we are able to absorb the effects of the coarse-graining
on the correlation function.  Moreover, in \refapp{peakbias} we show that
this continues to arbitrary powers of derivatives, and that the PBS bias
parameters can \emph{entirely} absorb the smoothing effect on $\xi_L$
[\refeq{xihpkPBS}].  Specifically, up to order $\nabla^4 \xi(r)$, we obtain
\ba
\xi_h(r) =\:& b_1^2 \xi(r) + 2 b_1 b_{\nabla^2\d} \nabla^2 \xi(r) \vs
& + \left[ (b_{\nabla^2\d})^2 + b_1 b_{\nabla^4\d}\right]\nabla^4 \xi(r).
\ea

There are two important implications of this result.  First, our approach, which
does not make any assumptions on the tracers themselves, generically predicts 
the existence of a bias with respect to $\nabla^2\d$, which
in $k$-space corresponds to a scale-dependent bias $\propto k^2$.  
One can interpret this as the statement that the tracer density is in general
not a truly local function of the matter density, but depends on the matter 
distribution within a finite region whose characteristic scale is given by
$\sqrt{b_{\nabla^2\d}}$.  

Second, the definition of the PBS bias parameter $b_{\nabla^2\d}$ [\refeq{bndef}] has
a clear physical interpretation: it corresponds to the response of the
tracer number density to a uniform shift in the curvature of the density
field (and shifts of higher derivatives in the general case, \refeq{alphatransgen}).  
Below we will show how this bias can be evaluated for an analytical
example, peaks of the density field.  A quantitative test on N-body simulations
will be the subject of future work.  

Beyond linear order in the matter correlation function,
one in principle has to expand the tracer density in
a multivariate bias expansion of $\nabla^{2n}\d_L$ ($n\geq0$).  
Although we have only shown that this expansion removes the $R_L$-dependence
contained in $\xi_L(r)$ at linear order in $\xi$, we expect this to be the 
case for higher orders as well.  In practice, the suppression of 
$b_{\nabla^{2n}\d} \nabla^{2n} \xi$ compared to $\xi$ ensures that one only needs to
keep a finite number of terms.  

\subsection{Connection with the peak model}

In the peak model, we identify large-scale structure tracers with discrete 
peaks of the density field above some threshold $\d_c$.  These distinct peaks 
constitute a point set (in contrast to the regions above threshold considered 
in \refsec{ex}).  For a Gaussian density field, it is possible to calculate the
two-point correlation of these peaks exactly 
\cite{desjacques:2008,desjacques/etal:2010}.  It is well 
known that peaks exhibit a scale-dependent bias in Fourier 
space $\propto k^2$ which is equivalent to a bias with respect to $\nabla^2 \d$
\cite{matsubara:1999,desjacques:2008,desjacques/etal:2010}.   
In this section, we show how the 
approach outlined in the previous section relates to this model.  

As in \refsec{ex}, we smooth the density field on a scale $R_*$ and define 
the local significance $\nu(\vx) = \d(\vx)/\s_*$.  Note that the smoothing scale 
here is physical, unlike the fictitious coarse-graining scale adopted in the 
renormalization approach described above.  
Apart from $R_*$, which is usually identified with the Lagrangian radius of the 
halos considered and is irrelevant for the discussion here, the peak model 
by itself does not involve any other coarse-graining scale.  
We begin by outlining the derivation of the mean number density of peaks, which 
follows App.~A of \cite{bardeen/etal:1986}, but will use the slightly different 
notation of \cite{desjacques:2008,desjacques/etal:2010}.  

The location $\vx_p$ of a peak is defined through constraints on the overdensity
$\d$ (or equivalently $\nu$), its gradient $\bm{\eta} \equiv \vn\delta$, and its 
Hessian $\zeta_{ij} \equiv \partial_i \partial_j \delta$:
\ba
\nu(\vx_p) >\:& \nu_c \s_* \vs
\bm{\eta}(\vx_p) =\:& 0 \vs
\lambda_1(\vx),\,\lambda_2(\vx),&\,\lambda_3(\vx) > 0,
\ea
where $\nu_c = \d_c/\s_*$ is the scaled threshold, and 
$\lambda_i$ are the eigenvalues of $\zeta_{ij}$.  Let $\v{V}$ denote
the 10-component vector consisting of $\nu, \bm{\eta}$, and the six
independent components of $\zeta_{ij}$.  The differential (in terms of $\nu$) 
mean number density of peaks is then given by
\ba
\bar n_{\rm pk}(\nu_c)\propto\:& \int \d_D(\nu-\nu_c) \d_D(\bm{\eta}) \Theta(\lambda_1)\Theta(\lambda_2)
\Theta(\lambda_3) \vs
& \qquad\times e^{-Q(\v{V})} d^{10}\v{V} \vs
Q(\v{V}) =\:& \frac12 \v{V}^T M^{-1} \v{V},
\ea
where $M$ is the covariance matrix of $\v{V}$.  The most difficult part
of the calculation is deriving the integration region and measure of $\v{V}$,
but we will not need to deal with this explicitly as we are only interested
in how $\bar n_{\rm pk}$ transforms under a change in $\ln\rhob$ and under
the transformation \refeq{curvtransf}.  Following \citet{bardeen/etal:1986} 
we introduce new variables $u,y,z$ ($u$ is their $x$), where in particular
\be
u \equiv -\frac{{\rm Tr}\, \zeta_{ij}}{\s_2}
= -\frac{\nabla^2 \d}{\s_2},\quad \s_2^2 = \<(\nabla^2\d)^2\>,
\ee
corresponds to minus the curvature of the density field scaled to unit variance.  
The log-likelihood $Q$ then becomes
\be
2 Q = \nu^2 + \frac{(u-u_*)^2}{1-\gamma^2} + 2Q_+,
\label{eq:LLQ}
\ee
where 
\be
u_* = \gamma\nu,\quad\mbox{and}\quad \gamma = \< \nu u\>
\ee
is a scaled spectral moment quantifying the correlation between $\nu$ and $u$.  
$2Q_+$ is the log-likelihood of $y$, $z$, $\bm{\eta}$, and the other 
components of $\zeta_{ij}$, and is independent of $\nu$ and $u$.  

\citet{bardeen/etal:1986} then find for the mean differential peak number 
density
\begin{equation}
\bar n_{\rm pk} = \frac{1}{(2\pi)^2 R_1^3}\, G_0(\gamma, \gamma \nu_c) e^{-\nu_c^2/2}\;,
\label{eq:npeaks}
\end{equation}
where $R_1$ is the characteristic radius of a peak [Eq.~(19) in \cite{desjacques/etal:2010}], and 
\begin{equation}
G_0(\gamma,u_*)=
\int_0^\infty\!\!du\, f(u) \frac{\exp\left(-\frac{(u-u_*)^2}{2(1-\gamma^2)}\right)}
{\sqrt{2\pi(1-\gamma^2)}}.
\end{equation}
Here, $f(u)$ is a function encoding the integral over $y$ and $z$ which
accounts for the asphericity of the peak profile \cite{bardeen/etal:1986}.  

Going back to \refeq{LLQ}, we see that adding a uniform density component corresponds
to a change in the threshold as described in \refsec{ex}, whereas the
other variables $u,y,z$ are not affected.  Thus, the density bias parameters 
are given by
\be
b_N = \frac{(-1)^N}{\bar n_{\rm pk}} \frac{\partial^N\bar n_{\rm pk}}{\partial(\delta_c)^N},
\label{eq:bNuniv}
\ee
where $\delta_c$ is the height of the density threshold.  These PBS bias 
parameters precisely
agree with the coefficients of $\xi(r)$ and $[\xi(r)]^2$ derived from an 
explicit computation of the peak 2-point 
correlation function \cite{desjacques/etal:2010}.  
The relation \refeq{bNuniv} was already pointed out in \cite{mo/jing/white:1997}.  

More interesting in this context is the derivation of the bias with respect
to $\nabla^2\d$.  We consider the transformation \refeq{curvtransf} in a
region of finite size $\L$, so that the effect on the density and $\nu$
is negligible if $\alpha \ll 1$.  Thus, all components of the vector
$\v{V}$ are unaffected with the exception of $u$.  Instead of following
a Gaussian of mean zero and variance of 1, $u$ is now Gaussian-distributed 
around (recall the minus sign in the definition of $u$)
\be
\<u\>_\alpha = -\frac{\alpha}{\s_2 \L^2},
\ee
with unit variance.  
Thus, the log-likelihood \refeq{LLQ} changes to
\be
2 Q(\alpha) = \nu^2 + \frac1{1-\gamma^2}\left(u+\frac{\alpha}{\s_2 \L^2} - u_*\right)^2 + 2Q_+.
\ee
Clearly, this is equivalent to changing
\be
u_* \to u_* - \frac{\alpha}{\s_2 \L^2}.
\ee
The remainder of the calculation of $\bar n_{\rm pk}$ follows through as before,
since it is independent of the distribution of $u$.  We thus 
obtain for the renormalized curvature bias [\refeq{bndef}] in the peak model
\ba
b_{\nabla^2\d} =\:& \frac{\L^2}{\bar n_{\rm pk}}\frac{\partial \bar n_{\rm pk}}{\partial \alpha}\Big|_{\alpha=0}
= - \frac{1}{\bar n_{\rm pk} \s_2}\frac{\partial \bar n_{\rm pk}}{\partial u_*}\Big|_{u_*=\gamma\nu_c}\vs
=\:& - \frac{1}{G_0(\gamma,\gamma\nu_c) \s_2}\frac{\partial G_0(\gamma,u_*)}{\partial u_*}\Big|_{u_*=\gamma\nu_c} \vs
=\:& -(G_0\s_2)^{-1}\int_0^\infty\!\!du\,
\left(\frac{u-\gamma\nu_c}{1-\gamma^2}\right)\,
f(u)\frac{e^{-\frac{(u-\gamma\nu_c)^2}{2(1-\gamma^2)}}}{\sqrt{2\pi(1-\gamma^2)}} 
\vs
=\:& -\frac{1}{\s_2}\left(\frac{\bar u - \gamma\nu_c}{1-\gamma^2}\right),
\label{eq:b01pk}
\ea
where $\bar u$ is the mean \emph{peak} curvature (i.e. the integral of $u$ times
the integrand of $G_0$, normalized by $G_0$ and evaluated at $\gamma\nu_c$). \refeq{b01pk} is precisely the 
scale-dependent bias parameter $b_{01}$ found in \cite{desjacques:2008,desjacques/etal:2010},
except for a minus sign which arises from the fact that these authors defined
$b_{01}$ as the linear bias associated with $\sigma_2 u = -\nabla^2\d$.  
The biases with respect to higher powers of $\nabla^2\d$ are obtained
by generalizing \refeq{b01pk} to
\be
b_{(\nabla^{2}\d)^N} = \frac{(-1)^N}{G_0(\gamma,u_*) \s_2^N}
\frac{\partial^N G_0(\gamma,\gamma\nu_c)}{\partial u_*^N}\Big|_{u_*=\gamma\nu_c}.
\label{eq:b0Npk}
\ee
In particular, we find for the PBS bias with respect $(\nabla^2\d)^2$
\ba
b_{(\nabla^{2}\d)^2} =\:& (G_0\s_2^2)^{-1}\int_0^\infty\!\!du\,
\frac{(u-\gamma\nu_c)^2 - (1-\gamma^2)}{(1-\gamma^2)^2}\,
f(u)\vs
& \qquad\qquad\times\frac{e^{-\frac{(u-\gamma\nu_c)^2}{2(1-\gamma^2)}}}{\sqrt{2\pi(1-\gamma^2)}} \vs
=\:& \frac1{\s_2^2} \left[\frac{\<(u - \gamma\nu_c)^2\>_{\rm pk}}{(1-\gamma^2)^2}- 
\frac1{1-\gamma^2} \right],
\ea
in agreement with $b_{02}$ as derived in \cite{desjacques:2013}.  Note that
by construction, the peak model does not predict any biasing with respect to 
higher than second derivatives of the density field ($b_{\nabla^{2n}\d} = 0$ for $n > 1$).  

\citet{desjacques/etal:2010} were also able to derive the scale-dependent bias 
$b_{01} = -b_{\nabla^2\d}$ from a peak-background split calculation.  However, 
they employed a conditional mass function, i.e. the number density of peaks
given a (spherical) overdensity on a much larger scale $R_B \gg R_*$, whereas 
here we derived all PBS bias parameters from the unconditional mass function 
(this is also the approach taken by \cite{desjacques:2013}).   
Nevertheless, all these treatments are based on a similar reasoning, i.e. 
a long-wavelength perturbation shifts the mean curvature (in the case of 
\cite{desjacques/etal:2010}, the shift is correlated with the density), and
lead to the same final result.  
Note that the peak-background split approach can be generalized to derive 
full expressions for the peak correlation functions \cite{desjacques:2013},  
which also includes dependencies on more general rotational invariants, such
as $(\vec\nabla\d)^2$ and $[(\partial_i\partial_j -\d_{ij}\nabla^2/3)\d]^2$.  
These are present in the clustering of peaks, even though it is not necessary
to introduce them in order to cure the $R_L$-dependence of $\xi_h$ induced
by smoothing, as we have seen here and in \refapp{peakbias}.  

It is important to note that the peak model is fundamentally different from 
local bias expansions in the sense that it only involves a physical smoothing 
scale $R_*$, and no further coarse-graining.  
Nevertheless, the term $\propto b_{01}\nabla^2\delta$ in $\xi_{\rm pk}$ is 
equivalent to the generic $b_{\nabla^2 \d}$ term in $\xi_h$. Namely, 
$b_{01}\nabla^2\delta$ restores the contrast of the baryon acoustic oscillation, 
otherwise smeared out by the filtering in $b_{10}^2\xi_*$ \cite{desjacques:2008}.  
Moreover, in general $b_{01}$ can be greater than $\sim R_*^2$ so that the contrast 
of the BAO in the peak correlation $\xi_{\rm pk}(r)$ can even be enhanced relative 
to that in the {\it unsmoothed} mass correlation $\xi(r)$ 
(see Fig.~5 in \cite{desjacques:2008}). 
However, subsequent gravitational evolution suppresses most of this scale-dependence 
(expected to be at the few percent level at the  time of collapse 
\cite{desjacques/etal:2010}).  
We expect this to be a general feature of a $\nabla^2\d$ bias specified in 
Lagrangian space.

\section{Non-Gaussian case}
\label{sec:NG}

We now return to the case of a tracer which can be sufficiently well described
by density bias, i.e. we neglect the curvature bias corrections, but consider
the case of initial conditions that are non-Gaussian.  Specifically, we will 
first focus
on the case of local primordial non-Gaussianity, which has been shown to lead
to a large modification of clustering on large scales;  general shapes of
non-Gaussianity will be considered in \refsec{nonloc}.    

The derivation in \refsec{PBS} can be generalized to a general
non-Gaussian density field, in which \refeq{xihPBS} formally retains 
its validity (see \refapp{general}).  However, one can easily show that in general the 
resulting tracer correlation function depends on the
coarse-graining scale $R_L$.  At lowest order the tracer auto-correlation
becomes
\be
\xi_h(r) = b_1^2 \xi_L(r) + b_1 b_2 \< \d_L(1) \d_L^2(2)\> + \O(\d_L^4),
\label{eq:xihNGloc}
\ee
the second term being the leading non-Gaussian correction.  
For primordial non-Gaussianity of the local type, and
in the limit $r \gg R_L$, the second correlator is given by (see \refeq{corrNGsq1})
\be
\<\d_L(1) \d_L^2(2) \> = 4\fNL \s_L^2 \xi_{\phi\d}(r),
\ee
where $\xi_{\phi\d}$ is the cross-correlation between the matter and
primordial Bardeen potential $\phi$.  Given the appearance of $\s_L^2$,
\refeq{xihNGloc} is strongly $R_L$-dependent.  This indicates that the 
description of the tracer density as a function of the 
matter density $\d_L$ alone is insufficient even on large scales in the
non-Gaussian case.  

Instead, we need to include a dependence of the
tracer density on the amplitude of small-scale fluctuations.  This
dependence is present regardless of the nature of the initial conditions;  
however, only in the non-Gaussian case are there
large-scale modulations of the small-scale fluctuations, due to mode
coupling, whereas in the Gaussian case we were able to neglect the
small-scale fluctuations in the large-scale description.  
In general,
one would imagine that the abundance of tracers depends on the amplitude
of small-scale fluctuations on a range of scales.  However, for simplicity
we will parametrize the dependence through the variance of the density
field on a single scale $R_*$.  In the local
model, which we will focus on here, this is sufficient in any case as
all small-scale fluctuations are rescaled equally (in the large-scale limit),
so that the value of $R_*$ becomes irrelevant for the final result.  

While we focus on primordial non-Gaussianity of the local type here, the
extension to other types of non-Gaussianity is straightforward (see \refsec{nonloc}).  
Furthermore, we only rely on the description of the density field in terms
of $N$-point functions, with the 3-point function being the lowest order
non-Gaussian contribution which we focus on here.  That is, we do not
rely on a fictitious Gaussian field from which the non-Gaussian field
is constructed.  This is different than the approach taken in 
\cite{giannantonio/porciani:2010,schmidt/kamionkowski:2010,scoccimarro/etal:2012}, where the separation of scales is typically applied in the fictitious Gaussian field, and an application in the physical non-Gaussian potential is not straightforward to implement \cite{scoccimarro/etal:2012}.  

We first define the small-scale density field as the local fluctuations around
the coarse-grained field $\d_L$:
\ba
\d_s(\vx) \equiv\:& \d_*(\vx) - \d_L(\vx) \label{eq:dsdef}\\
=\:& \int d^3\vy [W_*(\vx-\vy) - W_L(\vx-\vy)] \d(\vy) \vs
=\:& \int \frac{d^3\vk}{(2\pi)^3} \tilde W_s(k) \d(\vk)
e^{i\vk \cdot \vx}, \vs
\tilde W_s(k) =\:& \tilde W_*(k) - \tilde W_L(k).
\ea
In \reffig{sketch}, $\d_s$ is illustrated by the thin black line in the
lower part of the figure.  
Note that as $k\to 0$, $\tilde W_s(k) \propto k^2$, i.e. the long-wavelength modes are filtered
out as desired.  This implies that the cross-correlation between $\d_s(\vx)$
and the density field $\d_R(\vx)$ smoothed on some scale $R$ goes to
zero as $R$ becomes much larger than $R_L$:
\ba
\< \d_s(\vx) \d_R(\vx)\> =\:& \int \frac{d^3 k}{(2\pi)^3} \tilde W_s(k) \tilde W_R(k) P(k) \vs
\stackrel{R\gg R_L}{\to}\:& 0,
\label{eq:decorr}
\ea
and similarly for $\< \d_s(\vx_1) \d_L(\vx_2)\>$ if $|\vx_1-\vx_2| \gg R_L$.  
We further use the notation
\be
\s_s^2 \equiv \<\d_s^2\> = 
\int \frac{d^3 k}{(2\pi)^3} |\tilde W_s(k)|^2 P(k). 
\label{eq:sigmas2}
\ee
We quantify the dependence of the tracer abundance on the amplitude of 
small-scale fluctuations through 
\ba
y_*(\vx) \equiv\:& \frac12 \left(\frac{\d_s^2(\vx)}{\s_s^2} - 1\right) \; ,
\label{eq:ydef}
\ea
where the subscript $*$ refers to the smoothing scale $R_*$, $\<y_*\> =0$, 
and the factor of $1/2$ is included to obtain expressions which conform to 
standard convention later on.  
In the Gaussian case, $\xi_s(r) \to 0$ for $r \gg R_L$, so that the 
small-scale density field and $y_*$ in particular have no large-scale 
correlations.  
In the non-Gaussian case however, $y_*$ is in general correlated with 
long-wavelength perturbations.  Note that $\<\d_s(1)\d_L(2)\>$ vanishes
by construction on large scales [\refeq{decorr}], so that it is natural to start the
expansion with the leading term $\d_s^2$.    

We now generalize \refeq{Fp} to explicitly include the dependence on $y_*$,
\be
\nhat_h(\vx) = \Fh(\d_L(\vx), y_*(\vx);\vx).
\label{eq:FpNG}
\ee
Although our approach here is formally similar to the bivariate
local expansion in $\d_L$ and $\phi_L$ adopted in
\cite{mcdonald:2008,giannantonio/porciani:2010,baldauf/etal:2011},
there is somewhat of a conceptual difference in that we expand $\nhat_h$ 
purely in terms of properties of the matter distribution.  The effect of 
non-Gaussianity, and the fact that it derives from a potential $\phi$, 
only enter through the expressions for the correlators between $\d_L$
and $y_*$ here.  The nature of non-Gaussianity thus decouples from the 
description of the tracers (which only know about the matter density field)
in this approach.  

We can now repeat the derivation of \refsec{PBS}, including this
additional dependence.  All arguments about the residual scatter
from the deterministic relation $\nhat_h(\vx) = \nhat_h[\d_L(\vx),y_*(\vx)]$ 
and its negligible correlation with long-wavelength perturbations
made in \refsec{PBS} also apply here.  In fact, the dependence of
$\nhat_h(\vx)$ on $y_*(\vx)$ is a source of uncorrelated scatter
in the Gaussian case which becomes correlated with long-wavelength
perturbations in the non-Gaussian case.  This is another way of seeing
why we need to introduce the dependence on $y_*$ explicitly when dealing
with large-scale non-Gaussianity.  
Taking the expectation value of \refeq{FpNG}, we obtain
\ba
\<\nhat_h\> &\:= \<\Fh(0)\> \sum_{n,m} \frac{c_{nm}}{n!m!} 
\< \d_L^n y_*^m \>,\label{eq:npavgNG}
\ea
where we have defined bivariate ``bare'' bias parameters through
\be
c_{nm} \equiv \frac{1}{\<\Fh(0)\>} \left\<\frac{\partial^{n+m} \Fh}{\partial\d_L^n\partial y_*^m}\Big|_{\d_L=0, y_* = 0}\right\>.
\ee
We then need expressions for the various cross-correlations of 
$\d_L$ and $y_*$.  
In the following, we will restrict ourselves to the leading order 
terms, as the general expansion becomes lengthy.  

\subsection{Primordial non-Gaussianity of the local type}
\label{sec:localNG}

We will consider a density field derived from a 
Bardeen potential with non-Gaussianity of the local type.  We will
restrict our treatment to leading order in the non-linearity parameter
$\fNL$.  At this order, the only relevant $N$-point function is the
bispectrum,
\ba
B(\vk_1,\vk_2,\vk_3) =\:& \M(k_1)\M(k_2)\M(k_3) B_\phi(\vk_1,\vk_2,\vk_3)
\vs
B_\phi(\vk_1,\vk_2,\vk_3) =\:& 2\fNL [P_\phi(k_1)P_\phi(k_2) + (2~{\rm cyclic})].
\ea
Here,
\be
\M(k) = \frac{2}{3} \frac{k^2 T(k) g(z)}{\Omega_{m} H_0^2 (1+z)}
\ee
is the relation in Fourier space between the density and the Bardeen potential
$\phi$,
\be
\delta(\vk,z) = \M(k)\phi(\vk),
\ee
where $T(k)$ is the matter transfer function normalized to unity as $k\to 0$, 
and $g(z)$ is the linear growth rate of the gravitational potential normalized 
to unity during the matter dominated epoch.  Further, we define
$\M_L(k) = \M(k) \tilde W_L(k)$, $\M_s(k) = \M(k) \tilde W_s(k)$, and so on.  
We can then derive the leading contributions in the large-scale limit.  
As shown in \refapp{corrderiv},
\ba
\<\d_L(1) \d_L^2(2)\>  =\:&
\int \frac{d^3 k}{(2\pi)^3} e^{i\vk\cdot\vr} \M_L(k) 
\int \frac{d^3 k_1}{(2\pi)^3} \int \frac{d^3 k_2}{(2\pi)^3} \vs
&\times \M_L(k_1) \M_L(k_2) 
\< \phi_{\vk}\phi_{\vk_1}\phi_{\vk_2}\> \vs
=\:& 4 \fNL \s_L^2 \xi_{\phi\d,L}(r),
\label{eq:corrNGsq1}
\ea
where $\xi_{\phi\d,L}$ is the cross-correlation function between the
density coarse-grained on scale $R_L$ and the Bardeen potential $\phi$, i.e.
\be
\xi_{\phi\d,L}(r) = \int \frac{d^3 k}{(2\pi)^3} e^{i\vk\cdot\vr} \tilde W_L(k)
\int \frac{d^3 k_1}{(2\pi)^3} \< \d(\vk) \phi(\vk_1)\>.
\label{eq:xiphid}
\ee
In deriving \refeq{corrNGsq1}, we have expanded to lowest order in $k/k_1$ 
(``squeezed limit'' of the bispectrum), with the next higher
order being suppressed by $(k/k_1)^2$ in this limit.  We will discuss
this approximation in \refsec{disc}.  

Similarly, at leading order in $\fNL$ (see \refapp{corrderiv}),
\ba
& \<\d_L(1) y_*(2)\> = \frac12 \Big\< \d_L(1) \frac{\d_s^2(2)}{\s_s^2} \Big\> \vs
& = \frac1{2\s_s^2} \int \frac{d^3 k}{(2\pi)^3} e^{i\vk\cdot\vr} \M_L(k) 
\int \frac{d^3 k_1}{(2\pi)^3} \int \frac{d^3 k_2}{(2\pi)^3} \vs
&\qquad\qquad\times \M_s(k_1) \M_s(k_2)
\< \phi_{\vk}\phi_{\vk_1}\phi_{\vk_2}\> \vs
& = 2 \fNL \xi_{\phi\d,L}(r).
\label{eq:corrNGsq2}
\ea
This result can also be derived by using the well-known property of
local primordial non-Gaussianity that, 
in the squeezed limit, the local variance of the density field is rescaled
by $\phi(\vx)$, 
\be
\left.\left<\d_s^2(\vx)\right>\right|_{\phi(\vx)}
= \sigma_s^2 \left[1 + 4 \fNL \phi(\vx)\right] + \O(\fNL^2),
\label{eq:ds2NG}
\ee
and hence
\ba
\left.\left<y_*(\vx)\right>\right|_{\phi(\vx)}
=\:& 2 \fNL \phi(\vx) + \O(\fNL^2),
\ea
which immediately leads to \refeq{corrNGsq2}.  
Note that the correlators involving $y_*$ are independent of
the scale $R_*$ for local non-Gaussianity, so that the choice of
$R_*$ is arbitrary in this case.  We will see below how this changes for
other types of non-Gaussianity.  Note that \refeq{corrNGsq2} implies
that $y_*$ is of order $\phi$, i.e. linear in potential perturbations
 --- in contrast to the naive expectation that it is
of order $\d^2$.  Finally, we note that
$\< y_*(1) y_*(2)\>$ is $\O(\fNL^2)$ and hence not included
in the following.  


\subsection{Correlations}

We define the estimator for the correlation function through \refeq{xihat}
as before.  The expectation value then becomes
\ba
& \<\hat\xi_h(r)\> = \frac1{\N^2} 
\vs
& \times \sum_{n,m,n',m'=0}^\infty \!\!
\frac{c_{nm} c_{n'm'}}{n! m! n'! m'!} 
 \<\d_L^n(1)y_*^m(1)\d_L^{n'}(2)
y_*^{m'}(2)\>  - 1 ,
\label{eq:xih2NG}
\ea
where `1' and `2' stand for two arbitrary locations separated by a 
distance $r$, and we have redefined
\be
\N \equiv \sum_{n,m=0}^\infty \frac{c_{nm}}{n!m!} \<\d_L^n y_*^m\>.
\label{eq:NNG}
\ee
Similarly, we obtain the expectation value of the tracer-matter 
cross-correlation,
\ba
\<\hat\xi_{hm}(r)\> &\:= \frac1{\N} \sum_{n,m; n+m > 0}^\infty \frac{c_{nm}}{n!m!} 
\<\d_L^n(1) y_*^m(1)\d_L(2)\>. \label{eq:xihmNG}
\ea
Again, these expressions involve the ``bare'' bias parameters $c_{nm}$,
and the mixed moments of $\d_L,\:y_*$ which contain disconnected
pieces.  In the Gaussian case, \refeq{decorr} implies the absence
of any connected correlators involving $\d_L$ and $y_*$.   The powers of 
$y_*$ then only
add zero-lag pieces to the previous result \refeq{xih2}, which are
absorbed by corresponding terms in $\N$.  Thus, the final result
\refeq{xihPBSG} does not change in the Gaussian case if we include the 
dependence on $y_*$.  

Defining for convenience
\be
f(\vx) = \sum_{n,m=0}^\infty \frac{c_{nm}}{n!m!} \d_L(\vx) y_*(\vx) - 1,
\ee
we have 
\be
\xi_h(r) = \frac1{\N^2} \left[ \<f(1)f(2)\> - \< f \>^2\right],
\ee
where $\<f\> = \O(\d^2)$.  
In the following, we will expand $\xi_h$
to order $\d^4$, and simultaneously to linear order in $\fNL$ (as long as 
there are no other
sources of non-Gaussianity, going to $\O(\d^4)$ is also sufficient
to retain \emph{all} terms linear in $\fNL$).  Through
the latter restriction, we avoid a large number of quadratic and higher
order terms in $y_*$.   We have
\be
\N = 1 + \<f\> = 1 + \frac{c_{20}}2 \sigma_L^2 + c_{11} \<\d_L y_*\> + \O(\d^3),  
\ee
and
\be
\<f\>^2 = \frac{c_{20}^2}4 \sigma_L^4 + c_{11} c_{20} \<\d_L y_*\> \sigma_L^2 + \O(\d^5).
\ee
Hence, $\xi_h$ becomes 
\ba
\xi_h = \:& \frac1{\N^2} \Big\{\Big\langle\left( c_{10}\d_L + c_{01} y_*
+c_{11}\d_Ly_*
+ \frac{c_{20}}2 \d_L^2 + \frac{c_{30}}6 \d_L^3 \right)_1 \vs
& \left( c_{10}\d_L + c_{01} y_*
+c_{11}\d_Ly_*
+ \frac{c_{20}}2 \d_L^2 + \frac{c_{30}}6 \d_L^3 \right)_2
\Big\rangle \vs
& - \frac{c_{20}^2}4 \sigma_L^4 - c_{11} c_{20} \<\d_L y_*\> \sigma_L^2
\Big\}\vs
=\:& \frac1{\N^2} \Big[ c_{10}^2 \<\d_L(1)\d_L(2)\> + 2 c_{10} c_{01} \<\d_L(1) y_*(2)\> \vs
& + c_{10} c_{20} \< \d_L(1) \d_L^2(2)\> + c_{10} c_{30} \sigma_L^2 \<\d_L(1)\d_L(2)\> \vs
& + c_{01} c_{30} \sigma_L^2 \<\d_L(1)y_*(2)\> \vs
&  + 2 c_{11} c_{20} \<\d_L(1)y_*(2)\> \<\d_L(1)\d_L(2)\>\vs
& + 2\frac{c_{20}^2}4 \<\d_L(1)\d_L(2)\>^2 \Big] + \O(\d^5)
\label{eq:xihNG3}
\ea
where we have used the symmetry under interchange of locations 1 and 2,
and
\ba
\<y_*(1) \d_L^3(2)\> =\:& 3 \<y_*(1) \d_L(2)\> \sigma_L^2
+ \O(\fNL^2) \vs
\<\d_L(1)y_*(1)\d_L^2(2)\> =\:& 2 \<\d_L(1)y_*(2)\> \<\d_L(1)\d_L(2)\> \vs
& + \<\d_Ly_*\> \sigma_L^2 + \O(\fNL^2).
\label{eq:corrNG4th}
\ea
Perhaps somewhat surprisingly at first, we have to keep these terms whereas terms
such as $\<\d_L^2(1)y_*(2)\>,\:\<\d_L(1)y_*(1) \d_L(2)\>$ are higher order in $\fNL$ and thus
dropped.  This is simply because the latter terms do not have disconnected 
contributions.  

Note that all completely disconnected terms, i.e. terms that asymptote to
a constant as $r\to\infty$, have canceled as expected.  We now use the relations derived in
\refsec{localNG}.  Using \refeqs{corrNGsq1}{corrNGsq2}, we obtain
\ba
\xi_h(r) =\:& \frac1{\N^2} \Big[ \left( c_{10}^2  + c_{10} c_{30} \sigma_L^2\right) \xi_L(r) + \frac{c_{20}^2}{2} \xi_L(r)^2 \vs
& + \left(2 c_{10} c_{01} + c_{01} c_{30} \sigma_L^2 + 2 c_{10} c_{20} \sigma_L^2 \right) 2 \fNL \xi_{\phi\d,L}(r) \vs
& + 2 c_{11} c_{20} 2 \fNL \xi_{\phi\d,L}(r) \xi_L(r) \Big] .
\label{eq:xihNGsq}
\ea

\subsection{Bivariate PBS bias parameters}
\label{sec:PBSbiasNG}

In analogy to \refsec{PBSbias}, we would like to introduce a physically
motivated
bias parameter which quantifies the response of the tracer number density
to a change in the amplitude of small-scale fluctuations, 
without making reference to any
coarse-graining on the scale $R_L$.  The simplest way to parametrize such
a dependence
is to rescale all perturbations by a factor of $1+\eps$ from their fiducial
value, where $\eps$ is an infinitesimal parameter.  For example, for
a given realization of initial conditions of an N-body simulation, one
can obtain a realization with a different power spectrum normalization by
rescaling the initial density perturbations by $(1+\eps)$.\footnote{Of course,
if one initializes using a second-order density field, then the second
order part needs to be rescaled by $(1+\eps)^2$.}  Clearly, the
variance of the density field on some scale $R$, $\sigma_R^2$, is then rescaled 
to $(1+\eps)^2 \sigma_R^2$.    
Note that this means that the scaled cumulants $\<\d_*^n\>_c/\sigma_*^n$
are invariant, whereas the primordial non-Gaussianity parameter 
$\fNL \sim B_\Phi/P_\Phi^2$, if non-zero, scales as $(1+\eps)^{-1}$ 
under this transformation.  Specifically, under this rescaling
$\d_L$ and $y_*$ transform as
\ba
\d_L(\vx) \to\:& (1 + \eps) \d_L(\vx) \vs
y_*(\vx) \to\:& y_*(\vx) + \left(\eps + \frac{\eps^2}{2}\right) \frac{\d_s^2(\vx)}{\s_s^2} 
\;.
\label{eq:itransf}
\ea
Note that the parameter $\s_s^2$ in the definition of $y_*$ is just
a constant normalization, and does not change under the $\eps$-transformation.  
This is in analogy to keeping $\rhob$ fixed in the $D$-transformation in 
\refsec{PBSbias}.  

We can then define a set of \emph{bivariate PBS bias parameters} $b_{NM}$
by generalizing \refeq{bN} to
\be
b_{NM} \equiv \frac{1}{\<\nhat_h\>_{D=0,\eps=0}} \frac{\partial^{N+M} \<\nhat_h\>_{D,\eps}}{\partial D^N \partial \eps^M}\bigg|_{D=0,\eps=0}.
\label{eq:bNM}
\ee
These parameters can be understood as follows.  Given infinite volume,
the average tracer number density is a deterministic function of
the mean matter density $\rhob$ and the amplitude of the fluctuations
(parametrized, e.g., through the RMS of the density field on some
scale, $\sigma_*$).  $b_{NM}$ then denotes the $N+M$-th joint derivative of this function
with respect to $\ln \rhob$ and $\sigma_*$ (more precisely, $\eps$)
at some fiducial values of $\rhob$ and $\sigma_*$.  
Clearly, the parameters $b_{NM}$ are independent of the coarse-graining
scale $R_L$.    

As before, our next task is to derive the relation between $b_{NM}$ and
$c_{nm}$.  We have from \refeq{npavgNG},
\ba
\<\nhat_h\>(D,\eps) & = \<\Fh(0)\> \sum_{n,m=0}^{\infty} \frac{c_{nm}}{n!m!} \\
\times & \bigg\< \left[(1+\eps)\d_L+D\right]^n 
\left [ y_* + \left(\eps + \frac{\eps^2}{2}\right) \frac{\d_s^2}{\s_s^2}\right]^m \bigg\>.\nonumber
\ea
We thus have
\be
b_{N0} = b_N.
\ee
In particular,
\ba
b_{10} = \frac1{\N} \left( c_{10} + \frac{c_{30}}2 \sigma_L^2 + \O(\d_L^3)\right).
\ea
Further,
\ba
b_{01} =\:& \frac1{\N} \sum_{n,m} \frac{c_{nm}}{n!m!} 
\left(n \< \d_L^n y_*^m\> + m \< \d_L^n (1+2y_*) y_*^{m-1}\>\right)
\vs
=\:& \frac1{\N} \left( c_{01} 
+ c_{20} \sigma_L^2 + c_{11} \<\d_L y_*\>
+ \frac{c_{30}}2 \left<\delta_L^3\right>
+ \O(\d^4) \right).
\ea
We can now express the correlation function of tracers at this order,
\refeq{xihNG3}, in terms of the PBS bias parameters.  In fact,
if we are able to reach the analogous result to the Gaussian case,
i.e. that the tracer correlation function is a sum over PBS bias parameters
multiplying no-zero-lag correlators, we only need to keep terms up to
order $\d^2$ in $b_{NM}$, since they always multiply a correlator
of at least order $\d^2$.  Note that when extending the treatment
to higher order in $\fNL$, it is necessary to take into account that
$y_*$ transforms nonlinearly with $\eps$ [\refeq{itransf}].  This means that
the bias coefficient multiplying correlators containing say $y_*^2(1)$ will
not simply be $b_{N2}$, but involve a linear combination of $b_{N1}$ and $b_{N2}$.

Let us thus write all mixed ``no-zero-lag'' terms with the appropriate 
$b_{NM}$ in front, at order $\d^4, \fNL$.  We obtain
\ba
\xi_h(r) =\:& b_{10}^2 \xi_L(r) + \frac{b_{20}^2}{2} \xi_L^2(r)
+ 2 b_{10} b_{01} \<\d_L(1) y_*(2)\> \vs
& +    b_{20} b_{11} \< \d_L(1)y_*(1)\d_L^2(2)\>_{\rm nzl} \vs
& + \O(\d^5, \fNL^2).
\label{eq:xihPBSNG}
\ea
Here we have used the fact that at this order, $\<\d_L(1)\d_L^2(2)\>$,
$\<\d_L^2(1)y_*(2)\>$, $\<\d_L(1)\d_L^3(2)\>$, and $\<y_*(1)\d_L^3(2)\>$
have no \emph{no-zero-lag} pieces.  Note also that 
$\<\d_L(1) y_*(2)\> = \<\d_L(1) y_*(2)\>_{\rm nzl}$.  Plugging in the expressions for $b_{NM}$
at the relevant order, we obtain
\ba
\xi_h(r) =\:& \frac{1}{\N^2} \Big\{\left(c_{10}^2 + c_{10} c_{30} \sigma_L^2 \right) \xi_L(r) + \frac{c_{20}^2}{2} \xi_L^2(r) \vs
& + 2 \left( c_{10} c_{01} + \frac{c_{30}}2 c_{01} \sigma_L^2 + c_{10} c_{20} \sigma_L^2 \right) 2\fNL \xi_{\phi\d,L}(r) \vs 
& + 2 c_{20} c_{11} 2\fNL \xi_{\phi\d,L}(r) \xi_L(r) \vs 
& \Big\} + \O(\d^5, \fNL^2).
\ea
We easily see that this agrees identically with \refeq{xihNGsq}.  Thus,
the bivariate PBS parameters which we have defined in a coarse graining scale-independent
way absorb all coarse graining-scale dependent terms in the ``bare'' bias parameter
expansion \refeq{xih2NG}, in particular the term $c_1 c_2 \<\d_L(1) \d_L^2(2)\>$.  We expect this to hold to any
order in the bare bias parameter expansion, although a proof is beyond the
scope of this paper.    

Thus, the introduction of the bivariate bias parameters \refeq{bNM} and the
resulting expression \refeq{xihPBSNG}
achieved exactly what we had wanted.  In particular, the leading effect
of local primordial non-Gaussianity is quantified by $b_{01}$, the response
of the mean number density of tracers to a rescaling of the amplitude of
initial fluctuations.  The term $c_1 c_2 \<\d_L(1) \d_L^2(2)\>$ on the
other hand is seen as an artifact of the bare bias expansion which is absorbed 
in the renormalized parameter $b_{01}$.  
Apart from the clear physical interpretation, this reordering of the
perturbative expansion is also manifestly convergent:  higher order
terms are guaranteed to be suppressed by powers of $\xi_L(r)$ and $\fNL \xi_{\phi\d,L}(r)$,
which only need to be small on the scale of observation $r$ for the
perturbative expansion to be valid.  

This also remedies a worrying issue with the local bias expansion
in the non-Gaussian case:  evaluation of \refeq{xihPBS} shows that 
higher order terms (``loop corrections'') become comparable to or larger than
the leading order expression $b_1^2 \xi_L(r)$ on sufficiently large
scales, which would indicate a breakdown of the perturbative expansion.  
The bivariate expansion on the other hand leads to an expansion in which 
higher order terms are 
consistently suppressed [\refeq{xihPBSNG}], i.e. all dominating
terms are actually lowest order (``tree-level'').  For
sufficiently large values of $\fNL$, one might need to include higher order
terms in that parameter.  Nevertheless, the expansion will remain
convergent.  

\subsection{Non-local non-Gaussianity}
\label{sec:nonloc}

We now consider the generalization of the results of the last
section to arbitrary quadratic non-Gaussianity, i.e. non-Gaussianity
that is described to leading order by a 3-point function.  
The correlators that are relevant for the tracer two-point correlation
in the non-Gaussian case, \refeqs{corrNGsq1}{corrNGsq2}, are determined
by the behavior of the bispectrum in the squeezed limit, corresponding to
triangle configurations where one side
is much smaller than the other two.  For scale-invariant bispectra,
we can write the bispectrum in this limit as
\be
B_\phi(\vk_l, \vk_s, -\vk_l-\vk_s) \stackrel{k_l\ll k_s}{=}
A \left(\frac{k_l}{k_s}\right)^\alpha P_\phi(k_l) P_\phi(k_s),
\label{eq:BphiG}
\ee
with $A,\:\alpha$ being constants (more general shapes can be constructed by
linear superposition of bispectra with different $A_i,\:\alpha_i$).  
Local, folded, and equilateral
shapes correspond to $\alpha = 0,\:1$, and 2, respectively.  
\refeqs{corrNGsq1}{corrNGsq2} then generalize to
\ba
\<\d_L(1) \d_L^2(2)\>  =\:&
\int \frac{d^3 k}{(2\pi)^3} e^{i\vk\cdot\vr} \M_L(k) 
\int \frac{d^3 k_1}{(2\pi)^3} \vs
& \qquad\times \M_L^2(k_1) A k^\alpha P_\phi(k) k_1^{-\alpha} P_\phi(k_1) \vs
=\:& A \s_{-\alpha,L}^2 \xi_{\phi_\alpha\d,L}(r)
\label{eq:corrNGsqG1} \\
\<\d_L(1) y_*(2)\> =\:& \frac12 \Big\< \d_L(1) \frac{\d_s^2(2)}{\s_s^2} \Big\> \vs
=\:& \frac1{2 \s_s^2} \int \frac{d^3 k}{(2\pi)^3} e^{i\vk\cdot\vr} \M_L(k) 
\int \frac{d^3 k_1}{(2\pi)^3} \vs
& \qquad\times \M_s^2(k_1) A k^\alpha P_\phi(k) k_1^{-\alpha} P_\phi(k_1) \vs
=\:& A \frac{\s_{-\alpha,s}^2}{2 \s_s^2} \xi_{\phi_\alpha\d,L}(r),
\label{eq:corrNGsqG2}
\ea
where we have defined the general spectral moment
\be
\s_{n,X}^2 \equiv \int\frac{d^3k}{(2\pi)^3} k^n P(k) |\tilde W_X(k)|^2,
\ee
and the correlation between a non-local function of $\phi$ and the
density field,
\be
\xi_{\phi_\alpha\d,L}(r) \equiv \int \frac{d^3k}{(2\pi)^3} k^\alpha \M(k)
P_\phi(k) 
\tilde W_L(k).
\ee
Again, \refeqs{corrNGsqG1}{corrNGsqG2} are valid at leading order in the
squeezed limit ($k \ll k_1$, with corrections going as $(k/k_1)^2$).  
Inserting these expressions into \refeq{xihNG3}, and using \refeq{corrNG4th}
we obtain up to $\O(\d^5, \fNL^2)$
\ba
&\xi_h(r) = \frac1{\N^2} \Bigg[ \left( c_{10}^2  + c_{10} c_{30} \sigma_L^2\right) \xi_L(r) + \frac{c_{20}^2}{2} \xi_L(r)^2 \vs
& + \left(2 c_{10} c_{01} \frac{\s_{-\alpha,s}^2}{2 \s_s^2} + c_{01} c_{30} \frac{\s_{-\alpha,s}^2}{2 \s_s^2} \sigma_L^2 + c_{10} c_{20} \sigma_{-\alpha,L}^2 \right) \vs
& \quad\times A \xi_{\phi_\alpha\d,L}(r) \vs
& + 2 c_{11} c_{20} \frac{\s_{-\alpha,s}^2}{2 \s_s^2} A \xi_{\phi_\alpha\d,L}(r) \xi_L(r) \Bigg] .
\label{eq:xihNGsqG}
\ea
Inspection shows that the bivariate PBS parameters defined in \refsec{PBSbiasNG}
cannot absorb the $R_L$-dependent term from $\<\d_L(1)\d_L^2(2)\>$.  This
goes back to the fact that in the presence of a bispectrum of the form
\refeq{BphiG}, the small-scale perturbations are not rescaled uniformly,
but rather in a scale-dependent way:  the squeezed-limit result
\refeq{ds2NG} generalizes to 
\be
\left.\left<\d_s^2(\vx) \right>\right|_{\phi_\alpha(\vx)}
= \sigma_s^2 + A \s_{-\alpha,s}^2 \phi_\alpha(\vx),
\label{eq:ds2NGG}
\ee
where 
\be
\phi_\alpha(\vx) = \int \frac{d^3k}{(2\pi)^3} k^\alpha \phi(\vk) e^{i\vk\vx},
\label{eq:phialpha}
\ee
and hence
\ba
\left.\left<y_*(\vx)\right>\right|_{\phi_\alpha(\vx)}
=\:& A\frac{\s_{-\alpha,s}^2}{2 \s_s^2} \phi_\alpha(\vx).
\label{eq:lambdaG}
\ea
Thus, the transformation of the density field following \refeq{itransf}
is not the relevant one any more.  Instead, we need to rescale
the density field through
\be
\d(\vk) \to \left(1 + \eps k^{-\alpha} \right) \d(\vk),
\label{eq:itransfG}
\ee
so that
\ba
\d_L(\vx) \to\:& \d_L(\vx) + \eps\: \d_{-\alpha,L}(\vx) \vs
y_*(\vx) \to\:& y_*(\vx) + \frac{\eps}{\s_s^2} \d_s(\vx) \d_{-\alpha,s}(\vx)
+ \frac{\eps^2}{2\s_s^2} \d_{-\alpha,s}^2(\vx)
\; ,
\ea
where, in analogy to \refeq{phialpha},
\be
\d_{-\alpha,X}(\vx) \equiv \int \frac{d^3k}{(2\pi)^3} k^{-\alpha} \tilde W_X(k) \d(\vk)\,e^{i\vk\vx} .
\label{eq:dalpha} 
\ee
Note that $\<\d_X \d_{-\alpha,X}\> = \s_{-\alpha,X}^2$, and
$\<\d_{-\alpha,X}^2\> = \s_{-2\alpha,X}^2$.  
We will continue to assume that the tracer density depends on the
small-scale density field only through the variance on some scale $R_*$,
parametrized through $y_*$.  We again define $b_{NM}$ through \refeq{bNM}, 
but with the transformation \refeq{itransfG}, so that these bivariate
bias parameters will in general be different from those in \refsec{PBSbiasNG}.  
As before, our next task is to derive the relation between $b_{NM}$ and
$c_{nm}$.  
We have from \refeq{npavgNG},
\ba
\<\nhat_h\>(D,\eps) =\:& \<\Fh(0)\> \sum_{n,m} \frac{c_{nm}}{n!m!} \vs
& \times \Bigg\< \Big[(\d_L + \eps\, \d_{-\alpha,L})\d_L+D\Big]^n \vs
& \hspace*{0.7cm} \times
\left [y_* + \frac{\eps}{\s_s^2} \d_s \d_{-\alpha,s} + \frac{\eps^2}{2\s_s^2} \d_{-\alpha,s}^2 \right]^m \Bigg\>.\nonumber
\ea
We obtain
\ba
b_{01} =\:& \frac1{\N} \sum_{n,m} \frac{c_{nm}}{n!m!} \vs
&\quad\times \left(n \< \d_{-\alpha,L} \d_L^{n-1} y_*^m\> + \frac{m}{\s_s^2} 
\< \d_L^n  \d_s \d_{-\alpha,s} y_*^{m-1}\Big\>\right)
\vs
=\:& \frac1{\N} \left( c_{01} \frac{\s_{-\alpha,s}^2}{\s_s^2}
+ c_{20} \sigma_{-\alpha,L}^2 + \O(\d^3, \fNL^2) \right).
\ea
As in the case of local non-Gaussianity, we now write all mixed ``no-zero-lag'' 
correlators with the appropriate $b_{NM}$ in front,
up to $\O(\d^5, \fNL^2)$.  Due to the factor of $\s_{-\alpha,s}^2/\s_s^2$ in
the transformation of $y_*$ under the scale-dependent rescaling
\refeq{itransfG} (at lowest order), we have to divide by that factor
when multiplying correlators involving $y_*$.  We obtain
\ba
\xi_h &(r) = b_{10}^2 \xi_L(r) + \frac{b_{20}^2}{2} \xi_L^2(r)
+ 2 b_{10} b_{01} \frac{\s_s^2}{\s_{-\alpha,s}^2} \<\d_L(1) y_*(2)\> \vs
& + b_{20} b_{11} \frac{\s_s^2}{\s_{-\alpha,s}^2} \< \d_L(1)y_*(1)\d_L^2(2)\>_{\rm nzl} \vs
 = &\: b_{10}^2 \xi_L(r) + \frac{b_{20}^2}{2} \xi_L^2(r)
+ b_{10} b_{01} A \xi_{\phi_\alpha\d,L}(r) \vs
& + b_{20} b_{11} A \xi_{\phi_\alpha\d,L}(r) \xi_L(r). \label{eq:xihPBSNGG} 
\ea
Note that the final result is explicitly independent of the scale $R_L$
(as long as $r$ is sufficiently large so that the smoothing effect on $\xi_L(r)$,
$\xi_{\phi_\alpha\d,L}(r)$ is negligible), whereas $\<\d_L(1)y_*(2)\>$
itself is not since it depends on the spectral moment $\s_s^2$
[\refeq{corrNGsqG2}], which in turn depends on $\s_L^2$ [\refeq{sigmas2}].  

Inserting the expressions for $b_{NM}$ at the relevant order, and using
\refeqs{corrNGsqG1}{corrNGsqG2}, we have
\ba
\xi_h(r) & = \frac{1}{\N^2} \bigg\{\left(c_{10}^2 + c_{10} c_{30} \sigma_L^2 \right) \xi_L(r) + \frac{c_{20}^2}{2} \xi_L^2(r) \vs
& + \left[2 c_{10} c_{01} + 2 c_{10}c_{20} \s_{-\alpha,L}^2 \frac{\s_s^2}{\s_{-\alpha,s}^2} 
+ c_{30} c_{01} \s_L^2\right] \vs
& \quad\times \<\d_L(1)y_*(2)\>  \vs
& + c_{20} c_{11} \< \d_L(1)y_*(1)\d_L^2(2)\>_{\rm nzl}  \bigg\} \vs
=\:& \frac{1}{\N^2} \bigg\{ \left(c_{10}^2 + c_{10} c_{30} \sigma_L^2 \right) \xi_L(r) + \frac{c_{20}^2}{2} \xi_L^2(r) \vs
& + \left[2 c_{10} c_{01} + c_{30} c_{01} \s_L^2 \right]\frac{\s_{-\alpha,s}^2}{2 \s_s^2} A \xi_{\phi_\alpha\delta,R(r)} \vs
& + c_{10}c_{20} \s_{-\alpha,L}^2  A \xi_{\phi_\alpha\delta,R(r)} \vs
& + 2 c_{20} c_{11} \xi_L(r) \frac{\s_{-\alpha,s}^2}{2\s_s^2} A \xi_{\phi_\alpha\d,L}(r)
\bigg\}.
\ea
This agrees exactly with \refeq{xihNGsqG}.  The key difference of the
expansion of $\xi_h$ in terms of renormalized bias parameters in the
case of non-local primordial non-Gaussianity, \refeq{xihPBSNGG}, 
from the corresponding result for
local non-Gaussianity \refeq{xihPBSNG} is that the bivariate bias
parameters are now defined with respect to the scale-dependent 
rescaling of the density field, \refeq{itransfG}, rather than a
scale-independent rescaling.  We find that it is sufficient
in the large-scale limit, even in the case of a non-Gaussianity of general 
shape,
to describe the coarse-grained tracer abundance as a function
of $\d_L(\vx)$ and $y_*(\vx)$ in order to absorb the 
dependence on the coarse-graining scale $R_L$ into the bivariate
PBS bias parameters.  However, the
actual definition of the renormalized bias parameters depends on the
shape of primordial non-Gaussianity, in particular the scaling with
$k_l/k_s$ in the squeezed limit. 
 
We can thus summarize our findings regarding the effect of 
a primordial bispectrum on the two-point correlations
of tracers (non-Gaussian scale-dependent bias) as follows:
\begin{itemize}
\item For \emph{local} primordial non-Gaussianity, it is sufficient to
include the dependence of the tracer density $\nhat_h$ on the local amplitude
of small-scale fluctuations $\d_s$ through the variance on some scale $R_*$.  
Furthermore, the scale $R_*$ (and whether the dependence on $\d_s$ is
actually through the variance on several scales) is irrelevant, as all 
perturbations $\d_s$ are rescaled uniformly.  
\item For \emph{non-local separable} bispectra as in \refeq{BphiG}, it is
still sufficient to parametrize the dependence of $\nhat_h$ on the
amplitude of small-scale fluctuations through the variance on a single scale
$R_*$.  However, the value of the scale $R_*$ now matters as $y_*$
is modulated by an amount that depends on $R_*$ [\refeq{lambdaG}].  In
particular, if the tracer number density were to depend on the variance
of $\d_s$ on several different scales, then the PBS bias parameter $b_{01}$
will be a linear combination of these different dependencies with
relative weights controlled by $\alpha$, i.e. the shape of the bispectrum.  
\item For \emph{non-separable} bispectra, the renormalization approach
we describe here is not able to remove the $R_L$-dependence in the
tracer correlation function.  However, such shapes can typically be
well approximated by a linear superposition of separable shapes (see e.g. \cite{smith/zaldarriaga:2006}),
which then allows the renormalization to proceed as described here.  
\end{itemize}
Thus, we find that in general, a given tracer will respond differently
to different shapes of primordial non-Gaussianity, i.e. $b_{01}$ (and
$b_{NM}$ in general with $M > 0$) depends on the tracer as well as the
shape of the primordial bispectrum.  In the following we will study this
in the context of simplified models of tracers.

\subsection{Universal mass functions}
\label{sec:univ}

We begin with a generalization of the universal mass function discussed in \refsec{buniv}.  We write the mean abundance of tracers as 
\ba
\bar n_h =\:& \bar n_h\left(\bar\rho,\sigma_*, J_*\right)\,,
\label{eq:nhsimple}
\ea
where the Jacobian $J_*$ is defined in \refeq{Js}.  That is, $\bar n_h$ is given as a function of the mean density of the Universe and the variance of the density field smoothed on a scale $R_*$, as well as its derivative with respect to scale.  
Under the generalized rescaling \refeq{itransfG}, $\s_*$ transforms to lowest order as
\be
\sigma_* \to \sigma_*\left[1 + \eps \frac{\sigma_{-\alpha,*}^2}{\s_*^2}\right],
\ee
while the Jacobian transforms as (see also \cite{long})
\ba
J_* \to\:& J_*
+ \eps \frac{\s_{-\alpha,*}^2}{\s_*^2} \left( \frac{d\ln \s_{-\alpha,*}^2}{d \ln R_*}
-  \frac{d\ln \s_*^2}{d \ln R_*} \right) \vs
=\:& J_* \left[ 1 +
2 \eps \frac{\s_{-\alpha,*}^2}{\s_*^2} \left( \frac{d\ln \s_{-\alpha,*}^2}{d \ln \s^2_*}
-  1 \right) \right].
\ea
Here we have used $d/d\ln R_* = 2J_* \: d/d\ln \s_*^2$.  Note that in the local case
where $\alpha=0$, the local Jacobian is not affected by long-wavelength modes.  
Using \refeq{nhsimple}, we can then derive the leading non-Gaussian bias 
through \refeq{bNM}:
\ba
b_{01} & = \frac1{\bar n_h} \left(\frac{\partial \bar n_h}{\partial\ln\sigma_*}
\frac{\partial\ln \s_*}{\partial\eps}
+ \frac{\partial \bar n_h}{\partial \ln J_*}
\frac{\partial\ln J_*}{\partial\eps}  \right) \vs
=\:& \left[\frac1{\bar n_h}\frac{\partial \bar n_h}{\partial\ln \sigma_*}
+ \frac1{\bar n_h}\frac{\partial \bar n_h}{\partial \ln J_*} 
2\left( \frac{d\ln \s_{-\alpha,*}^2}{d \ln \s^2_*} -  1 \right) \right]
\frac{\s_{-\alpha,*}^2}{\s_*^2} 
 \vs
=\:& \left[ b_{01}(\alpha=0) 
+ 2\left( \frac{d\ln \s_{-\alpha,*}^2}{d \ln \s^2_*}
-  1 \right) \right]\frac{\s_{-\alpha,*}^2}{\s_*^2}.
\ea
Here, $b_{01}(\alpha=0)$ is the PBS bias parameter quantifying the
effect of local primordial non-Gaussianity for a tracer following
\refeq{nhsimple}, and we have assumed that the tracer density scales linearly with
the Jacobian as expected physically.  For such tracers, the bias parameters quantifying
the response to general non-local non-Gaussianity (in the squeezed limit)
are thus directly related to those for local non-Gaussianity.  In particular, 
we recover the results of \cite{long}, who first
pointed out the contribution by the Jacobian $J_*$.   

We now specialize \refeq{nhsimple} to a ``truly'' universal mass function
[\refeq{univ1}],
\be
\bar n_h = \rhob\:f(\nu_c)\: J_*\,,\quad \nu_c\equiv \frac{\d_c}{\s_*}\,,
\label{eq:univ}
\ee
where $f(\nu_c)$ is in general an arbitrary function of $\nu_c$.  
The results relating $b_{01}(\alpha)$ to $b_{01}(\alpha=0)$ 
of course also hold in this case.  However, the 
specific form \refeq{univ} further allows us to connect $b_{01}(\alpha=0)$
to the linear PBS density bias:
\ba
b_{10} =\:& \frac1{\bar n_h} \frac{\partial \bar n_h}{\partial\ln \rhob}
= - \frac1{\s_*} \frac{df}{d\nu_c} \vs
b_{01}(\alpha=0) =\:& \frac1{\bar n_h} \frac{\partial \bar n_h}{\partial\ln \sigma_*}
= - \frac{\d_c}{\s_*} \frac{df}{d\nu_c} = \d_c b_{10}.
\label{eq:b01univ}
\ea
Note that here $b_{10}$ is the Lagrangian bias, which is why we have not
included the derivative with respect to $\ln\rhob$ of the $\rhob$ prefactor
in \refeq{univ} (see also \refsec{buniv}); again, the effect on $J_*$ vanishes for
$\alpha=0$.  This is the original relation
between the density bias parameter and the response to primordial
non-Gaussianity derived in \cite{dalal/etal:2008,slosar/etal:2008,giannantonio/porciani:2010}.  We point out that these results differ from those of \cite{matsubara:2012}, who considered the effect of primordial non-Gaussianity on tracers with local Lagrangian biasing.  There, the entire leading order effect of primordial non-Gaussianity is encoded in a scale-dependent $c_2(\vk_1,\vk_2)$.  Thus, a parametrization of $c_2(\vk_1,\vk_2)$ down to very small scales is necessary in order to predict the amplitude of the scale-dependent bias.  This is in contrast to the approach presented here, where one introduces a local dependence on the small-scale fluctuations which absorbs the term proportional to $c_2$ into a renormalized $b_{01}$, which is a single number.  As a result, the prediction of \cite{matsubara:2012} yields a departure from \refeq{b01univ} for universal mass functions which depends on the precise form of $f(\nu_c)$.  While \refeq{b01univ} has been both supported \cite{dalal/etal:2008,desjacques/seljak/iliev:2009,giannantonio/porciani:2010} and disputed \cite{grossi/etal:2009} by simulation results, these different predictions are clearly resolvable with sufficiently large simulations.  In particular, our prediction for the scale-dependent bias for a general tracer,
\be
b_{01} = \frac{1}{\<\nhat_h\>}\frac{\partial \<\nhat_h\>}{\partial\eps}\,,
\ee
which is independent of any assumptions on the mass function of the tracer, provides a rigorous test of our approach which can be applied to simulations.

\section{Summary \& Discussion}
\label{sec:disc}

We have shown that the expression of tracer correlations in terms of
$R_L$-independent renormalized bias parameters $b_N$ absorbs all zero-lag correlators present in the expansion of the tracer correlation function in terms of the bare ``scatter-plot'' bias parameters $c_n$.  We have shown this to all 
orders for an arbitrary density field.  While the proof only applies directly for the auto- and cross-correlation functions and pure density biasing,
we expect the result to hold in the case of higher $N$-point functions
and multivariate biasing as well (analogously to the resummed multipoint propagators of \cite{bernardeau/crocce/scoccimarro:2008,matsubara:2011}).  
Our key result is a rigorous definition of the renormalized bias parameters
in terms of derivatives of the mean number density of tracers with respect
to the background density (we call these ``peak-background split'' bias
parameters since their definition is closely related to the commonly adopted
definition of PBS biases \cite{kaiser:1984,cole/kaiser:1989,sheth/tormen:1999}).  
It is important to stress that this exact definition is entirely independent
of the nature of the tracer considered.  Therefore, it
provides a rigorous framework in which further assumptions for or modeling of
the bias parameters, for example from the excursion set, peak model, or 
halo occupation distribution, can be embedded. 

Our results go beyond previous work on renormalized bias parameters 
\cite{mcdonald:2006} in two ways:  first, we show that our result is valid to all orders;  second,
we rigorously connect the renormalized bias parameters with the peak-background
split.  We further show that the renormalized bias parameters in the
tracer auto-correlation and the tracer-matter cross-correlation agree to all
orders.  We
also expect this to be the case for higher $N$-point functions, although
this remains to be shown.   We can summarize this reasoning as in  
line (a) of \reftab{sum}:  the expression of tracer correlations in terms
of the bare biases $c_n$ is $R_L$-dependent at each order due to disconnected 
correlators (for example $c_2^2 \s_L^2$).  This $R_L$-dependence is then
resummed into $R_L$-independent bias parameters $b_N$ which are defined
with respect to a uniform increase in the matter density.  

The underlying assumption in this result is that the clustering of tracers
is entirely determined by their dependence on the local matter density.  
This is not expected to be a good assumption in general.  However,
our result provides another invaluable tool:  whenever
the renormalized expression in terms of no-zero-lag correlators
exhibits a residual dependence on $R_L$, we conclude that a biasing purely
in terms of matter density is not sufficient.  

\begin{table}[b!]
\centering
\begin{tabular}{c|c|c|c}
 & $R_L$-dependence & local & Transformation defining \\
 & & quantity & PBS bias parameter \\
\hline\hline
(a)\   & $c_2^2 \s_L^2 \xi_L(r)$ & $\delta_L$ & $\rho \to \rho + D\rhob$ \\
\hline
(b)\   & $\xi_L(r)$ & $\nabla^2\d_L$ & $\nabla^2\d \to \nabla^2\d + \alpha/\L^2$ \\
\hline
(c)\   & \  $c_1 c_2 \<\d_L(1)\d_L^2(2)\>$\   & $y_*$ & $\delta \to (1+\eps) \delta$ \\
\hline
\end{tabular}
\caption{Summary of renormalization procedures introduced to remove various
dependencies of tracer correlations on the coarse-graining scale $R_L$.
The variable $y_*$ is defined in \refeq{ydef}.
\label{tab:sum}}
\end{table}

We first encounter this in the case of the smoothed matter correlation function
$\xi_L(r)$, which depends on $R_L$ if $\xi(r)$ has structure on scales smaller
than $R_L$ (line (b) in \reftab{sum}).  
In this case, we are led to introduce bias parameters with
respect to the curvature (Laplacian) of the matter density field.  In 
Fourier space, this corresponds to a scale-dependent bias $\propto k^2$.  If we
further include bias parameters with respect to higher derivatives
of the density field, we can in fact \emph{entirely} absorb the effect of 
smoothing on $\xi(r)$ [\refapp{peakbias}].  Again, this is regardless of the nature of the tracer
and the shape of the matter correlation function.  Of course, for a smooth
correlation function, it is usually sufficient to keep only terms involving
the lowest few derivatives of the density field.  
The renormalized biases with respect to the curvature correspond
to derivatives of the mean tracer abundance with respect to a 
constant shift in the curvature of the density field (\reftab{sum}).  As an example, these
bias parameters are easily derived for peaks of a Gaussian density field 
from the results of \cite{bardeen/etal:1986}.  We show that the bias
parameters obtained in this way indeed match the scale-dependent biases
derived in the full, direct calculation of peak correlations \cite{desjacques:2008}.  
In this context, it is important to point out that the curing of
$R_L$-dependencies, such as that from a smoothing of the correlation function,
is a sufficient condition for having to introduce an additional dependence
of the tracer density on properties of the matter density field.  However, it
is not a necessary condition---specific tracers might also exhibit additional
dependencies not required by renormalization.  One example is peaks of the
matter density field, which also exhibit a dependence on quantities such as
$(\vec\nabla\d)^2$ \cite{desjacques:2013}.  Of course, it is straightforward to
include these additional dependencies in the formalism described here, by defining
renormalized PBS bias parameters through suitable transformations of the density
field.

In the case of a non-Gaussian density field, we find that the tracer 
correlation function for pure density biasing acquires a strong dependence
on $R_L$ if long-wavelength modes are coupled to short wavelength modes (line (c) in \reftab{sum}).  
The most well known example of this kind is primordial non-Gaussianity of 
the local type \cite{mcdonald:2008}.  In this case, we have to add a bias
parameter with respect to the amplitude (variance) of small-scale fluctuations.  
The renormalization procedure then absorbs the $R_L$-dependent terms
such as $c_1 c_2 \<\d_L(1) \d_L^2(2)\>$, and the resulting bivariate 
bias parameters are given
by the derivatives of the mean tracer density with respect to the background
density and (essentially) the amplitude of the initial power spectrum---both clearly
$R_L$-independent quantities.  Effectively, we obtain an expansion
closely related to that of \cite{giannantonio/porciani:2010}, although
we did not need to drop any terms or make approximations beyond the large-scale
limit (which allows us to evaluate the bispectrum in the squeezed limit).  

We also generalize the results to any form of primordial non-Gaussianity
given through a bispectrum of potential perturbations.  In fact, this 
provides a good example for how this renormalization approach pays off:  
we obtain a fully general and exact result (in the large-scale limit), in 
which the renormalized scale-dependent bias parameter depends on the precise 
shape of the non-Gaussianity as well as the nature of the tracer.  Assuming 
that the tracer abundance only depends on the variance of the small-scale
density field on a single scale, we can however relate
the scale-dependent bias parameter for an arbitrary general shape to that
for local non-Gaussianity.  Further, we can be more restrictive and
assume that the tracer follows a universal mass function.  In this case,
we can relate the scale-dependent bias parameter to the bias parameter
with respect to density (as in  \cite{dalal/etal:2008,slosar/etal:2008,giannantonio/porciani:2010}).  

The general procedure also carries over to primordial non-Gaussianity 
described by higher $N$-point functions.  For example, a non-zero trispectrum
which couples long- to short-wavelength modes will introduce a significant
$R_L$-dependence in the tracer correlation function through the term
$c_1 c_3\<\d_L(1) \d_L^3(2)\>$.  In order to remedy this, we need to
explicitly take into account the dependence of the tracer density on the
local skewness $\<\d_s^3\>$ of the density field, which then yields
a corresponding scale-dependent bias contribution (as shown in \cite{long,smith/ferraro/loverde:2012}) 
which absorbs the $R_L$-dependent terms.  

The main caveat to our results is that we have worked in Lagrangian
space throughout.  While we expect that the general approach will also
be applicable in Eulerian space, the effect of gravitational evolution
will in general introduce several further dependencies of the 
tracer density on the environment, for example velocity and tidal field
biases \cite{desjacques/sheth:2010,chan/scoccimarro/sheth:2012,baldauf/etal:2012}.  
We leave this for future work.  Further, we have neglected the effects of
supersonic relative motion between baryons and dark matter \cite{tseliakhovich/hirata:2010}, which are potentially important for low-mass tracers at high redshifts.  If relevant, this effect can be included through an additional bias with respect to the relative velocity squared \cite{dalal/etal:2010,yoo/etal:2011}.  Note that the statistical properties of this relative velocity are very well understood.

We have also only considered observables in real space.  The main 
obstacle in transforming to Fourier space is the issue of stochasticity
in the tracer density field and its scale dependence, which contributes
to correlations at all $k$ in Fourier space (although the contributions will
asymptote to a constant in the low-$k$ limit).  Thus, a well-defined model
for correlations on small scales is a necessary prerequisite for a rigorous 
understanding of Fourier-space correlations.    

Further, we have restricted the treatment here to two-point correlations
of tracers.  
The main reason for this is simplicity;  we expect no major obstacles in
generalizing the results to higher $N$-point functions  
in Lagrangian space, such as the tracer bispectrum with non-Gaussian initial conditions.  In order for this
to be useful however, non-Gaussianities from gravitational evolution
will also have to be included
\cite{2007PhRvD..76h3004S,jeong/komatsu:2009b,2009PhRvD..80l3002S,liguori/etal:2010,2010arXiv1011.1513B,sefusatti/etal:2012}.

Finally, in the case of primordial non-Gaussianity, we have only considered
linear terms in $\fNL$, and restricted to the large-scale limit where the 
bispectrum is evaluated at lowest order
in the ``squeezed-limit'' expansion.  The extension to higher powers of
$\fNL$ is straightforward.  The second approximation captures the main effects
on large scales, since the subleading term is suppressed by $k_l^2/k_s^2$, 
where $k_l \sim 1/r$ is the scale on which we measure
correlations, and $k_s \sim 1/R_*$ corresponds to the small-scale
fluctuations.  For example, in case of local non-Gaussianity, the
subleading term is expected to lead to a small approximately scale-independent bias.

These caveats notwithstanding, we hope these results provide the starting 
point for a rigorous treatment of biasing of general tracers in the context 
of cosmological perturbation theory.

\acknowledgments
VD and FS thank Marc Kamionkowski and Physics and Astronomy at Johns Hopkins
University for hospitality.  FS thanks Tobias Baldauf, Eiichiro Komatsu, Daan Meerburg, Roman Scoccimarro,
and Svetlin Tassev for discussions.   FS was supported by the National Aeronautics and Space Administration through Einstein Postdoctoral Fellowship Award Number PF2-130100 issued by the Chandra X-ray Observatory Center, which is operated by the Smithsonian Astrophysical Observatory for and on behalf of the National Aeronautics Space Administration under contract NAS8-03060.  
DJ is supported by DoE SC-0008108 and NASA NNX12AE86G.  
VD acknowledges support by the Swiss National Science Foundation.


\appendix

\section{Proof of \refeq{xihPBS} for a general density~field}
\label{app:general}

Let us denote as $\Pi_n$ the set of all partitions of the set
\be
\{\underbrace{1,\,1,\,\dots,\,1}_{n\:\rm times}\},
\ee
where the elements of the set are considered distinguishable.  
We call the elements $B$ of any given $\pi \in \Pi_n$ ``blocks'', with $\pi$ having
$|\pi|$ blocks where $|\pi|$ is the cardinality, or number of elements, of $\pi$.  Clearly, $|\pi| \leq n$, 
and the blocks of any partition in $\Pi_n$ satisfy
\be
\sum_{B\in \pi} |B| = n.
\ee
For example for $n=4$ there are four distinct partitions with one block $B_1$ with
$|B_1| = 1$ and one block $B_2$ with $|B_2|=3$.  
Then, the moment for an arbitrary
density field $\d$ is given in terms of the cumulants (connected correlators) by
\be
\< \d^n \> = \sum_{\pi\in \Pi_n} \prod_{B\in \pi} \< \d^{|B|}\>_c\;.
\label{eq:mom1}
\ee
For example, in this sum the trivial partition 
$\pi = \big\{\{1,\,1,\,\dots,\,1\} \big\}$ (with a single block $B$ with $|B|=n$)
corresponds to $\< \d^n\>_c$.  Note that since $\<\d\>=0$, any partition
where $|B|=1$ for any $B\in\pi$ yields a vanishing contribution.  

Similarly, let us denote as $\Pi_{n,m}$ the set of all partitions of
\be
\{\underbrace{1,\,1,\,\dots,\,1}_{n\:\rm times},\,
\underbrace{2,\,2,\,\dots,\,2}_{m\:\rm times}\}.
\ee
We can then write
\ba
\<\d_1^n \d_2^m\> = \:& \sum_{\rho\in \Pi_{n,m}} \prod_{B \in \rho}
\Big\< \prod_{a \in B} \d_a \Big\>_c
\vs
=\:& \sum_{\rho\in \Pi_{n,m}} \prod_{B \in \rho}
\< \d_1^{n_1(B)} \d_2^{n_2(B)} \>_c \;,
\label{eq:momgen}
\ea
where in the first line $a$ runs over the elements of the block $B$, and in the
second line we have defined as $n_1(B)$ the number of elements `1' in block $B$,
and correspondingly for $n_2(B)$ (so that $n_1(B)+n_2(B) = |B|$).  
This simplifies the result since the cumulants
are independent of the order of products of $\d_1$ and $\d_2$, and only depend on
the overall power of each.

Our goal is to reorder the sum in \refeq{momgen}.  We assign two further
numbers (non-negative integers to be precise) to each $\rho\in\Pi_{n,m}$:
\be
\bar{N}_1(\rho) = \sum_{B \in \rho}^{n_2(B)=0} n_1(B);\quad
\bar{N}_2(\rho) = \sum_{B \in \rho}^{n_1(B)=0} n_2(B)\;.
\ee
In other words, for a given partition $\rho$, $\bar{N}_1(\rho)$ counts the 
number of elements `1' that are in blocks
that \emph{only} contain `1', while $\bar{N}_2(\rho)$ counts the number of `2's that
are in blocks only containing `2'.  
These numbers are of course uniquely defined for each $\rho$.  Moreover,
$\bar{N}_1,\,\bar{N}_2$ define a partition of $\Pi_{n,m}$, i.e. each $\rho \in \Pi_{n,m}$
is member of one and only one subset of $\Pi_{n,m}$ defined as containing
all $\rho$ with a specific value of $\bar{N}_1$ and $\bar{N}_2$.  Equivalently,
the relation $\rho\sim \sigma$ defined for any $\rho,\sigma \in \Pi_{n,m}$ through
\be
\rho\sim \sigma \  \Leftrightarrow\   \bar{N}_1(\rho) = \bar{N}_1(\sigma) \wedge 
\bar{N}_2(\rho) = \bar{N}_2(\sigma)
\ee
is an equivalence relation on $\Pi_{n,m}$.  
We can then split the sum in \refeq{momgen} into sums over these 
disjoint subsets of $\Pi_{n,m}$:
\be
\<\d_1^n \d_2^m\> = \sum_{k=0}^n \sum_{l=0}^m \sum_{\rho \in \Pi_{n,m}}^{\bar{N}_1(\rho)=k;\: \bar{N}_2(\rho)=l} 
\prod_{B \in \rho}
\< \d_1^{n_1(B)} \d_2^{n_2(B)} \>_c \;.
\ee
Consider the sum over all partitions in one of these subsets:
\be
\sum_{\rho\in \Pi_{n,m}}^{\bar{N}_1(\rho)=k;\: \bar{N}_2(\rho)=l} 
\prod_{B \in \rho}
\< \d_1^{n_1(B)} \d_2^{n_2(B)} \>_c \;.
\label{eq:sumkl}
\ee
A partition $\rho \in \Pi_{n,m}$ can be thought of as one specific way
of distributing $n$ black balls and $m$ red balls into arbitrarily many (initially empty)
boxes.  These boxes correspond to the cumulants 
in \refeq{momgen} (of course empty boxes are trivial, because they yield 1 in the
product in \refeq{momgen}; boxes with only one ball lead to a zero
contribution).  The sum in \refeq{sumkl} runs over all possible ways of distributing
these balls that have exactly $k$ black balls which are in boxes
with only black balls, and $l$ red balls which are in boxes with only
red balls.  Correspondingly, the remaining $n-k$ black and $m-l$ red balls
are in boxes with \emph{both} black and red balls.  
There are  $(n\quad k)$ ways of selecting $k$ black balls out of $n$,
and $(m\quad l)$ ways for the red balls.  Given such a selection of
$k$ out of $n$ and $l$ out of $m$, the sum in \refeq{sumkl} thus runs over
all ways of partitioning $k$ black balls into boxes, $m$ red balls
into a different set of boxes, and finally $n-k$ black and $m-l$ red balls 
into a third set of boxes
such that each of these boxes contains at least one black and one red.  
Mathematically, we can write \refeq{sumkl} as
\begin{widetext}
\ba
\sum_{\rho\in\Pi_{n,m}}^{\bar{N}_1(\rho)=k;\: \bar{N}_2(\rho)=l} \prod_{B \in \rho}
\< \d_1^{n_1(B)} \d_2^{n_2(B)} \>_c =\:&
\binom{n}{k}
\left(\sum_{\pi_1 \in \Pi_k} \prod_{B_1 \in \pi_1} \<\d_1^{|B_1|}\>_c \right)
\binom{m}{l}
\left(\sum_{\pi_2 \in \Pi_l} \prod_{B_2 \in \pi_2} \<\d_2^{|B_2|}\>_c \right) \vs
&\times 
\sum_{\rho \in \Pi_{n-k,m-l}^{\rm nzl}}
\prod_{B\in \rho}
\< \d_1^{n_1(B)} \d_2^{n_2(B)}\>_c
\vs
=\:&
\binom{n}{k} \<\d_1^k\>
\binom{m}{l} \<\d_2^l\>
\sum_{\rho \in \Pi^{\rm nzl}_{n-k,m-l}}
\prod_{B\in \rho} \< \d_1^{n_1(B)} \d_2^{n_2(B)}\>_c\;.
\label{eq:momgen2a}
\ea
\end{widetext}
Here, we have used \refeq{mom1}, and defined the subset $\Pi^{\rm nzl}_{n,m}$ of 
the set of all partitions $\Pi_{n,m}$ through
\ba
\Pi^{\rm nzl}_{n,m} =\:& \Big\{ \rho \in \Pi_{n,m}:\  \bar{N}_1(\rho) = 0 = \bar{N}_2(\rho) \Big\} \vs
=\:& \Big\{ \rho \in \Pi_{n,m}:\  \forall B\in \rho\  n_1(B) > 0 \wedge n_2(B) > 0  \Big\}.\nonumber
\ea
In other words, $\Pi^{\rm nzl}_{n,m}$ contains all partitions in which each block has
at least one element `1' and at least one element `2'.  That is, for 
any $\rho \in \Pi^{\rm nzl}_{n,m}$, the product of correlators
\be
\prod_{B\in \rho} \< \d_1^{n_1(B)} \d_2^{n_2(B)} \>_c
\ee
\emph{does not contain any zero-lag pieces}.  It is then natural to define the 
no-zero-lag correlator for a general density field as 
\be
\< \d_1^n \d_2^m \>_{\rm nzl} \equiv
\sum_{\rho \in \Pi^{\rm nzl}_{n,m}}
\prod_{B\in \rho}
\< \d_1^{n_1(B)} \d_2^{n_2(B)}\>_c\;,
\label{eq:nzldef}
\ee
which reduces to \refeq{cnzlG} for a Gaussian density field.  

Finally, we can sum \refeq{momgen2a} over all values of $\bar{N}_1(\rho),\:\bar{N}_2(\rho)$ to
obtain
\be
\<\d_1^n \d_2^m\> = \sum_{k=0}^n\sum_{l=0}^m 
\binom{n}{k} \<\d_1^k\>
\binom{m}{l} \<\d_2^l\>
\<\d_1^{n-k} \d_2^{m-l}\>_{\rm nzl}\;.
\label{eq:momgen2}
\ee
We are now ready to prove \refeq{xihPBS} for general non-Gaussian matter density
fields.  Plugging the relation between $b_N$ and $c_n$, \refeq{bNcN}
into \refeq{xihPBS}, and relabeling $N\to n-N$, $M\to m-M$, yields
\ba
\<\hat \xi_h(r)\> =\:& \frac1{\N^2} \sum_{n,m=0}^\infty \frac{c_n c_m}{n! m!}
\sum_{N=0}^{n-1} \sum_{M=0}^{m-1}
\binom{n}{N} \binom{m}{M} \vs
& \hspace*{1.2cm} \times \<\d_L^N\> \<\d_L^M\> \<\d_L^{n-N}(1) \d_L^{m-M}(2)\>_{\rm nzl} \vs
=\:& \frac1{\N^2} \sum_{n,m=0}^\infty \frac{c_n c_m}{n! m!} 
\left[\<\d^n_l(1) \d^m_l(2)\>
- \<\d_L^n\> \<\d_L^m\> \right] \vs
=\:& \frac1{\N^2} \sum_{n,m=0}^\infty \frac{c_n c_m}{n! m!} 
\<\d^n_l(1) \d^m_l(2)\> - 1.
\ea
In the second line, we have used \refeq{momgen2} and subtracted out
the completely disconnected contribution $\<\d_L^n\> \<\d_L^m\>$
(which is not included in \refeq{xihPBS} as we start the sum from
$N=1$, $M=1$).  In the third line, we have used \refeq{N}.  This is
identical to \refeq{xih2}, and thus proves the relation \refeq{xihPBS} 
between PBS bias parameters and two-point correlations
for a general non-Gaussian density field at all orders.

\section{Curvature bias to higher order}
\label{app:peakbias}

Let us consider a general spherically symmetric filter function,
\be
W_L(\vx) = \frac1{4\pi R_L^3}\,f\left(\frac{|\vx|}{R_L}\right),
\label{eq:Wf}
\ee
where $f(y)$ is a dimensionless function defined on $[0,\infty)$.  
We have pulled out a factor of $(4\pi R_L^3)^{-1}$ for convenience.  
The Fourier transform of $W_L(\vx)$ can then be written as
\ba
\tilde W_L(k) =\:& \sum_{n=0}^\infty (k R_L)^{2n} \frac{(-1)^n}{(2n+1)!} f_n \vs
f_n \equiv\:& \int_0^{\infty} dy\: y^{2n+2} f(y).
\label{eq:Wtgen}
\ea 
The normalization constraint $\int d^3 \vx\: W_L(\vx) = 1$ is equivalent
to $f_0=1$.  Since $R_L$ is an arbitrary parameter, we can further choose one of the
$f_n$ with $n\geq 1$ to assume some desired value.  Specifically, in \refsec{peakbias} 
we have chosen $R_L$ so that $f_1 = 6$ and hence $W(k) = 1-k^2R_L^2 +\cdots$.    

The smoothed correlation function can then be written \emph{exactly} as
\ba
\xi_L(r) =\:& \sum_{n,m=0}^\infty \frac{f_n f_m R_L^{2(n+m)}}{(2n+1)!(2m+1)!} \nabla^{2(n+m)} \xi(r). \label{eq:xiRexp}
\ea
Note that the factor of $(-1)^{n+m}$ from the expansion of $\tilde W_L$
cancels with the $i^{2(n+m)}$ from converting powers of $k$ into derivatives.  

We can generalize the transformation \refeq{curvtransf} as follows:
\be
\d_{\alpha}(\vy) = \d(\vy) + \alpha \sum_{N=0}^\infty \frac{g_N}{\L^{2N}} |\vy|^{2N},
\label{eq:alphatransgen}
\ee
where $\alpha$ and $g_N$ are dimensionless parameters, and the case $N=0$
corresponds to the transformation used to derive the PBS density bias 
(with $g_0=1$ and $\alpha=D$, \refsec{PBSbias}).  Using the expansion of the filter function, we obtain
\be
\d_{L,\alpha}(\v{0}) = \d_L(\v{0}) + \alpha \sum_{N=0}^\infty \left(\frac{R_L}{\L}\right)^{2N} g_N f_N.
\ee
Further, using that
\ba
\nabla^2 |\vy|^{2N} =\:& (2N+1)2N |\vy|^{2N-2} \vs
\Rightarrow \nabla^{2n} |\vy|^{2N} =\:& \frac{(2N+1)!}{(2N-2n+1)!} |\vy|^{2(N-n)},
\ea
for $n \leq N$, we obtain in analogy to \refeq{xiRexp}
\ba
& \nabla^{2n} \d_{L,\alpha}(\v{0}) = \nabla^{2n} \d_L(\v{0}) \vs
& + \frac{\alpha}{\L^{2n}} 
\sum_{N=n}^{\infty} \left(\frac{R_L}{\L}\right)^{2(N-n)}\!\!\frac{(2N+1)!}{(2N-2n+1)!} g_N f_{N-n}.
\ea
The transformation \refeq{alphatransf} corresponds to $g_1 = 1/6$ with all
other $g_n$ equal to zero.  Hence,
\ba
\nabla^2 \d_{L,\alpha}(\v{0}) =\:& \nabla^2\d_L(\v{0}) + \frac{\alpha}{\L^2} 3! \frac16 \vs
=\:& \nabla^2\d_L(\v{0}) + \frac{\alpha}{\L^2} 
\ea
as intended (with all higher derivatives being unaffected).  
More generally, we can define $g_N = 1/(2N+1)!$ for a fixed $N$, with all 
other $g_n=0$, so that
\ba
& \nabla^{2n} \d_{L,\alpha^{(N)}}(\v{0}) = \nabla^{2n}\d_L(\v{0}) \vs
& + \frac{\alpha^{(N)}}{\L^{2n}} \left(\frac{R_L}{\L}\right)^{2(N-n)}\!\!
\frac{1}{(2N-2n+1)!} f_{N-n},
\ea
where $n\leq N$.  In particular,
\ba
\d_{L,\alpha^{(N)}}(\v{0}) =\:& \d_L(\v{0}) + \alpha^{(N)} 
\left(\frac{R_L}{\L}\right)^{2N}\frac{1}{(2N+1)!} f_N \vs
\nabla^{2N}\d_{L,\alpha^{(N)}}(\v{0}) =\:& \nabla^{2N} \d_L(\v{0}) + \frac{\alpha^{(N)}}{\L^{2N}}.
\ea
Now, if we write
\be
\hat n_h(\vx) = \Fh\left(\left\{ \nabla^{2n} \d_L(\vx)\right\}_{n=0}^{\infty};\vx\right),
\ee
we can define the generalized, bare linear curvature bias parameters (at linear order in
$\nabla^{2n}\d$) through
\be
c_{\nabla^{2n}\d} \equiv \frac1{\Fh(0)} \frac{\partial \Fh\left(\left\{ \nabla^{2m} \d_L(\vx)\right\}_{m=0}^{\infty};\vx\right)}{\partial (\nabla^{2n}\d)}\Big|_0.
\ee
Further, we can define PBS bias parameters as
\ba
b_{\nabla^{2N}\d} =\:& \frac{\L^{2N}}{\<\hat n_h\>}\frac{\partial\<\hat n_h\>_{\alpha^{(N)}}}{\partial\alpha^{(N)}}\Big|_{\alpha^{(N)}=0} \label{eq:bnabdef}\\
=\:& \L^{2N} \sum_{n=0}^N c_{\nabla^{2n}\d} \frac{\partial \nabla^{2n} \d_{L,\alpha^{(N)}}}{\partial \alpha^{(N)}} \vs
=\:& \sum_{n=0}^N c_{\nabla^{2n}\d} \frac{R_L^{2(N-n)}}{(2N-2n+1)!} f_{N-n},
\label{eq:bnabcnab}
\ea
where we have taken out powers of $\L$ to make the expression for the 
correlation function below simpler.  Note that the usual PBS density bias
is obtained as special case for $N=0$.  

In terms of the bare bias parameters, and to linear order in matter correlations,
we can write the tracer correlation function as
\ba
\xi_h(r) =\:& \sum_{n,m=0}^\infty c_{\nabla^{2n}\d} c_{\nabla^{2m}\d} 
\nabla^{2(n+m)} \xi_L(r) \vs
=\:& \sum_{n,m=0}^\infty c_{\nabla^{2n}\d} c_{\nabla^{2m}\d} \vs
& \!\!\! \times\sum_{\tilde N,\tilde M=0}^\infty \frac{f_{\tilde N} f_{\tilde M} R_L^{2(\tilde N+\tilde M)}}{(2\tilde N+1)!(2\tilde M+1)!}  
\nabla^{2(\tilde N+\tilde M+n+m)} \xi(r) 
\vs
=\:& \sum_{N,M=0}^{\infty} \sum_{n,m=0}^{N,M} c_{\nabla^{2n}\d} c_{\nabla^{2m}\d} \label{eq:corrpkexact}\\
& \!\!\! \times \frac{f_{N-n} f_{M-m} R_L^{2(N-n + M-m)}}{(2N - 2n + 1)!(2M-2m+1)!} 
\nabla^{2(N+M)} \xi(r).\nonumber
\ea
In the second line, we have shifted the sum by defining $N \equiv \tilde N + n$, 
$M \equiv \tilde M + m$, while in the third line we have reordered the sum over
$n,m$ and $N,M$.  

Now let us write down the expected result in terms of PBS bias parameters,
i.e. assuming that the parameters $b_{\nabla^{2N}\d}$ have absorbed all
dependencies on $R_L$:
\ba
\xi_h(r) =\:& \sum_{N,M=0}^{\infty} b_{\nabla^{2N}\d} b_{\nabla^{2M}\d} \nabla^{2(N+M)} \xi(r),
\label{eq:xihpkPBS}
\ea
where we have again restricted to linear order in matter correlations.  
Using \refeq{bnabcnab}, this yields
\ba
\xi_h(r) =\:& \sum_{N,M=0}^{\infty} \sum_{n,m=0}^{N,M} c_{\nabla^{2n}\d} c_{\nabla^{2m}\d} \vs
& \times \frac{R_L^{2(N+M-n-m)} f_{N-n} f_{M-m}}{(2N-2n+1)!(2M-2m+1)!}  
\nabla^{2(N+M)} \xi(r),\nonumber
\ea
which agrees with the exact result \refeq{corrpkexact}.  Thus, the PBS
bias parameters defined in \refeq{bnabdef} are indeed able to completely
absorb the effects of smoothing on the correlation function (at linear
order).  Moreover, if we assume that there is a characteristic scale $\L$
describing the dependence of the tracer density on the derivatives
of the density field, i.e. 
\be
b_{\nabla^{2N}\d} \sim \L^{2N},
\ee
then the expression \refeq{xihpkPBS} is an expansion in terms of 
\be
(\L^2 \nabla^2)^N \xi(r).
\ee
If the matter correlation function does not have significant structure on scales
below $\L$, then this quantity is progressively suppressed at higher $N$
and \refeq{xihpkPBS} is rapidly convergent.  

\section{Derivation of \refeqs{corrNGsq1}{corrNGsq2}}
\label{app:corrderiv}

In this appendix we derive the squeezed-limit expressions
\refeqs{corrNGsq1}{corrNGsq2}.  We begin with the correlator \refeq{corrNGsq1}:
\ba
 \<\d_L(1) \d_L^2(2)\>  =\:&
\int \frac{d^3 k}{(2\pi)^3} e^{i\vk\cdot\vr} \M_L(k) 
\int \frac{d^3 k_1}{(2\pi)^3} \int \frac{d^3 k_2}{(2\pi)^3} \vs
& \times
\M_L(k_1) \M_L(k_2) \< \phi_{\vk}\phi_{\vk_1}\phi_{\vk_2}\>.
\nonumber
\ea
Using the definition of the bispectrum,
\be
\< \phi_{\vk}\phi_{\vk_1}\phi_{\vk_2}\>
= (2\pi)^3 \d_D(\vk+\vk_1+\vk_2) B_\phi(\vk,\vk_1,\vk_2),
\ee
we have
\ba
 \<\d_L(1)\d_L^2(2)\> =\:&
 \int \frac{d^3 k}{(2\pi)^3} e^{i\vk\cdot\vr} \M_L(k) 
\int \frac{d^3 k_1}{(2\pi)^3} 
\M_L(k_1) \vs
& \times \M_L(|\vk+\vk_1|) 
B_\phi(k,k_1,|\vk+\vk_1|)  . 
\label{eq:diffterms}
\ea
We now expand the integrand in powers of $q = k/k_1$.  Further, we
define $\mu = \hat\vk\cdot\hat\vk_1$.  We then obtain for any
function $F(k)$
\ba
F(|\vk + \vk_1|) =\:& F(k_1)\bigg[ 1 + \left(q\mu + \frac12q^2[1-2\mu^2]\right) a(k_1) \vs
& \hspace*{1cm} + \frac12 q^2\mu^2 b(k_1) \bigg] + \O(q^3), \vs
a \equiv\:& \frac{\partial\ln F(k_1)}{\partial\ln k_1} \vs
b \equiv\:& \frac1{F(k_1)}\frac{\partial^2 F(k_1)}{\partial(\ln k_1)^2}
\ea
while
\ba
& B_\phi(k,k_1,|\vk+\vk_1) = 2\fNL \Big\{ P_\phi(k) \left[P_\phi(k_1) + P_\phi(|\vk+\vk_1|)\right] \vs
& \hspace*{4.2cm} + P_\phi(k_1) P_\phi(|\vk+\vk_1|) \Big\} \vs
=\:& 2\fNL P_\phi(k) P_\phi(k_1) 
\left[ 2 + (2q\mu + [1-2\mu^2]q^2) n_\phi \right] + \O(q^3), \nonumber
\ea
where $n_\phi = n_s-4$ and we have assumed a pure power-law $P_\phi(k)$ 
for simplicity.  Inserting these expressions into \refeq{diffterms}, we
see that the terms $\propto q\mu$ vanish once the integral over $\mu$
is performed.  Thus, we obtain
\ba
\<\d_L(1) \d_L^2(2)\> =\:&  4\fNL
\int \frac{d^3 k}{(2\pi)^3} e^{i\vk\cdot\vr} \M_L(k)  P_\phi(k) \vs
&\times \int \frac{d^3 k_1}{(2\pi)^3} \M_L^2(k_1) P_\phi(k_1)
\left [1 + \O(q^2)\right]. \nonumber
\ea
The $k_1$ integral yields $\s_L^2$, leading to \refeq{corrNGsq1} with corrections suppressed in the large-scale limit by $(k/k_1)^2 \sim (k R_L)^2$.  
We now turn to \refeq{corrNGsq2}.  Since
\be
\< \d_L(1) y_*(2) \> = \frac1{2\s_s^2} \<\d_L \d_s^2(2)\>,
\ee
this just differs from \refeq{corrNGsq1} through the prefactor and
the different filter function, and the above results immediately
lead to the third line of \refeq{corrNGsq2}.

\bibliography{PBSng}

\end{document}